\documentclass[preprint,eqsecnum,aps,nofootinbib]{revtex4}
\usepackage{amsfonts,amsmath,amssymb,amsthm}
\usepackage{latexsym}
\usepackage{bbm,bm}
\usepackage{graphicx}

\newcommand{\beq}{\begin{equation}}
\newcommand{\eeq}{\end{equation}}
\newcommand{\beqs}{\begin{eqnarray}}
\newcommand{\eeqs}{\end{eqnarray}}

\begin{document}

\title{Dynamics of Entanglement in Three Coupled Harmonic Oscillator System with Arbitrary Time-Dependent Frequency and Coupling Constants}

\author{DaeKil Park$^{1,2}$}

\affiliation{$^1$Department of Electronic Engineering, Kyungnam University, Changwon
                 631-701, Korea    \\
             $^2$Department of Physics, Kyungnam University, Changwon
                  631-701, Korea    
                      }

\begin{abstract}
The dynamics of mixedness and  entanglement is examined by solving the  time-dependent Schr\"{o}dinger equation for three coupled harmonic oscillator system with arbitrary time-dependent  frequency and coupling constants parameters. We assume that part of oscillators is inaccessible and remaining oscillators
accessible. We compute the dynamics of entanglement between inaccessible and accessible oscillators. 
In order to show the dynamics pictorially we introduce three quenched models. In the quenched models both mixedness and entanglement exhibit oscillatory 
behavior in time with multi-frequencies. It is shown that the mixedness for the case of one inaccessible oscillator is larger than that for the case of two
inaccessible oscillators in the most time interval. Contrary to the mixedness entanglement for the
case of one inaccessible oscillator is smaller  than that  for the case of two inaccessible oscillators in the most time interval. 
\end{abstract}

\maketitle

\section{Introduction}
The most peculiar and counterintuitive properties of quantum mechanics are superposition and entanglement\cite{schrodinger-35,text,horodecki09} of quantum states. In addition to their importance from a pure theoretical aspect, entanglement is known to play a crucial role in the quantum information 
processing  such as  quantum teleportation\cite{teleportation},
superdense coding\cite{superdense}, quantum cloning\cite{clon}, quantum cryptography\cite{cryptography,cryptography2}, and quantum
metrology\cite{metro17}. It is also quantum entanglement, which makes the quantum computer outperform the classical one\cite{qcreview,computer}.
Since quantum technology developed by quantum information processing attracts a considerable attention recently due to limitation of classical
technology, it is important to understand the various properties of entanglement.

In the theory of entanglement the most basic questions are how to detect and how to quantify it from given quantum states. For last two decades 
these questions have been explored mainly in the qubit system. The strategy to first question is to construct the entanglement witness operators and 
to explore their properties and applications\cite{detect}. Second question has been explored by constructing the various entanglement measures 
such as distillable entanglement\cite{benn96}, entanglement of formation \cite{benn96}, relative entropy of entanglement\cite{vedral-97-1,vedral-97-2},
three tangle\cite{ckw,ou07-1} {\it et cetera}.

In spite of construction of many entanglement measures the analytic computation of these measures is very difficult even in the qubit
system\footnote{However, it is possible to compute entanglement of formation for arbitrary two-qubit state\cite{woot-98}.} except very rare cases. 
In the real physical system where the quantum state is dependent on continuum variables computation of such measures is highly difficult or 
might be impossible. Frequently, thus, we use the von Neumann\cite{woot-98} and R\'{e}nyi entropies\cite{renyi96} to measure the bipartite entanglement
of continuum state. Furthermore, the entropies enable us to understand the Hawking-Bekenstein entropy\cite{bekenstein73,hawking76,hooft85,luca86,mark93,solo11} of black holes more deeply. They are also important to study on the 
quantum criticality\cite{eisert10,vidal03} and topological matters\cite{levin06,jiang12}. 

In this paper we will study on the dynamics of entanglement in the three coupled harmonic oscillator system  when frequency and coupling constant parameters
are arbitrary time-dependent.  The harmonic oscillator system is used in many branches of physics due to its mathematical simplicity. 
The analytical expression of von Neumann entropy was derived for a general real Gaussian density matrix in 
Ref. \cite{luca86} and it was generalized to massless scalar field in Ref. \cite{mark93}. Putting the scalar field system in the spherical box, the author 
in Ref. \cite{mark93} has shown that the total entropy of the system is proportional to surface area. This result gives some insight into a question why 
the Hawking-Bekenstein entropy of black hole is proportional to the area of the event horizon. Recently, the entanglement is computed in the coupled harmonic oscillator  system using a Schmidt decomposition\cite{maka17}. The von Neumann and R\'{e}nyi entropies are also explicitly computed in the similar system, called two site Bose-Hubbard model\cite{ghosh17}. More recently, the dynamics of entanglement and uncertainty is exactly derived in the two coupled 
harmonic oscillator system  when frequency and coupling constant parameters are arbitrary time-dependent\cite{park18}.

In this paper we assume as follows. Let us consider three coupled harmonic oscillators $A$, $B$, and $C$, whose  frequency and coupling constant parameters
are arbitrary time-dependent. Let us assume part of oscillator(s) is inaccessible. For example, part of oscillator(s) falls into black hole horizon and 
as a result, we can access only remaining ones. Under this situation we derive the time-dependence of entanglement between 
inaccessible and accessible oscillators analytically. As a by-product we also derive the time-dependence of mixedness, which is trace of square of reduced 
quantum state. If mixedness is one, this means the quantum state is pure. It it is zero, this means the quantum state is completely mixed. 

This paper is organized as follows. In next section the diagonalization of Hamiltonian is discussed briefly. In Sec. III we derive the solutions for
 time-dependent Schr\"{o}dinger equation (TDSE) explicitly in the coupled harmonic oscillator system. In Sec. IV we derive the time dependence of
 entanglement when $A$ and $B$ oscillators are inaccessible. The time dependence of mixedness for $C$ oscillator is also derived. In Sec. V 
 we derive the time dependence of entanglement when $A$ oscillator is inaccessible. The time dependence of mixedness for $(B, C)$-oscillator system is 
 also derived. In section VI we introduce three sudden quenched models, where the frequency and coupling constants are abruptly changed at $t=0$. Using the results of 
 previous sections we compare the dynamics of entanglement and mixedness when the inaccessible oscillator(s) is different. In Sec. VII a brief conclusion is
 given. In appendix A the quantities $\alpha_i$, $\beta_i$, and $\gamma_{ij}$, which appear in the reduced quantum state and  have long expressions, are explicitly summarized. 
 
\section{Diagonalization of Hamiltonian}
The Hamiltonian we will examine in this paper is
\begin{eqnarray}
\label{hamil-1}
&&H = \frac{1}{2} (p_1^2 + p_2^2 + p_3^2) + \frac{1}{2} \bigg[ K_0 (t) (x_1^2 + x_2^2 + x_3^2) + J_{12} (t) (x_1 - x_2)^2   \\   \nonumber
&&  \hspace{5.0cm} + J_{13} (t) (x_1 - x_3)^2 
+ J_{23} (t) (x_2 - x_3)^2  \bigg]
\end{eqnarray}
where $\{x_i, p_i \} \hspace{.1cm}  (i=1, 2, 3)$ are the canonical coordinates and momenta. We assume that the frequency parameter $K_0$ and coupling
constants $J_{ij}$ are arbitrarily time-dependent. The Hamiltonian can be written in a form
\begin{equation}
\label{hamil-2}
H = \frac{1}{2} \sum_{j=1}^3 p_j^2 + \frac{1}{2} \sum_{i, j=1}^3 x_i K_{ij} (t) x_j
\end{equation}
where
\begin{eqnarray}
\label{def-K}
K (t) = \left(                  \begin{array}{ccc}
             K_0 + J_{12} + J_{13}   &    -J_{12}     &    -J_{13}                              \\
             -J_{12}    &    K_0 + J_{12} + J_{23}    &    -J_{23}                              \\
             -J_{13}    &    -J_{23}     &  K_0 + J_{13} + J_{23}  
                                    \end{array}                             \right).
\end{eqnarray}

The eigenvalues of $K(t)$ are $\lambda_1 (t) = K_0$ and $\lambda_{\pm} (t) = K_0 + J_{12} + J_{13} + J_{23} \pm z$, where 
\begin{equation}
\label{def-z}
z (t) = \sqrt{J_{12}^2 + J_{13}^2 + J_{23}^2 - \left(J_{12} J_{13} + J_{12} J_{23} + J_{13} J_{23} \right)}.
\end{equation}
The corresponding normalized eigenvectors are 
\begin{eqnarray}
\label{eigen-1}
v_1 (t) = \frac{1}{\sqrt{3}}  \left(   \begin{array}{c} 1 \\ 1  \\ 1  \end{array}   \right)      \hspace{2.0cm}
v_{\pm} (t) = A_{\pm}  \left(    \begin{array}{c} -J_{12} + J_{23} \mp z  \\  J_{12} - J_{13} \pm z  \\  J_{13} - J_{23} \end{array}   \right)    
\end{eqnarray}
where 
\begin{equation}
\label{normal-1}
A_{\pm} (t) = \frac{1}{J_{13} - J_{23}}  \left( \frac{2 z \pm \left(J_{13} + J_{23} - 2 J_{12} \right)}{6 z} \right)^{1/2}.
\end{equation}
Since $K(t)$ is symmetric, $v_j \hspace{.1cm}  (j=1, \pm)$ are orthonormal to each other. It is worthwhile noting 
\begin{equation}
\label{useful-1}
A_+^2 A_-^2 = \frac{1}{12 z^2 (J_{13} - J_{23})^2},
\end{equation}
which is frequently used later. Thus, $K (t)$ can be diagonalized as $K (t) = U^{t} (t) K_D (t) U (t)$, where 
\begin{eqnarray}
\label{diagonal-1}
U (t) = \left(        \begin{array}{ccc}
                  1 / \sqrt{3}    &    1 / \sqrt{3}    &    1 / \sqrt{3}                                                           \\
                  A_+ (-J_{12} + J_{23} - z)    &    A_+ (J_{12} - J_{13} + z)    & A_+ (J_{13} - J_{23})             \\
                  A_- (-J_{12} + J_{23} + z)    &    A_- (J_{12} - J_{13} - z)    & A_- (J_{13} - J_{23}) 
                        \end{array}                                                \right)
\end{eqnarray}
and $K_D (t) = \mbox{diag} (\lambda_1, \lambda_+, \lambda_- )$. 

Now, we introduce new coordinates 
\begin{eqnarray}
\label{change-1}
 \left(   \begin{array}{c} y_1 \\ y_+  \\ y_-  \end{array}   \right)  = U (t)
  \left(   \begin{array}{c} x_1 \\ x_2  \\ x_3  \end{array}   \right).
\end{eqnarray}
In terms of the new coordinates the Hamiltonian (\ref{hamil-2}) can be diagonalized in a form
\begin{equation}
\label{diag-hamil}
H = \frac{1}{2} \left[ \pi_1^2 + \omega_1^2 (t) y_1^2 \right] + \frac{1}{2} \left[ \pi_+^2 + \omega_+^2 (t) y_+^2 \right] 
+ \frac{1}{2} \left[ \pi_-^2 + \omega_-^2 (t) y_-^2 \right] 
\end{equation}
where $\pi_j$ are conjugate momenta of $y_j$ and $\omega_j (t) = \sqrt{\lambda_j} \hspace{.2cm} (j = 1, \pm)$.

\section{Solutions of TDSE}
Consider a Hamiltonian of single harmonic oscillator with arbitrarily time-dependent frequency
\begin{equation}
\label{hamil-3}
H_0 = \frac{p^2}{2} + \frac{1} {2} \omega^2 (t) x^2.
\end{equation}
The TDSE of this system was exactly solved in Ref. \cite{lewis68,lohe09}. The linearly independent solutions $\psi_n (x, t) \hspace{.1cm} (n=0, 1, \cdots)$ are expressed in a form
\begin{equation}
\label{TDSE-1}
\psi_n (x, t) = e^{-i E_n \tau(t)} e^{\frac{i}{2} \left( \frac{\dot{b}}{b} \right) x^2} \phi_n \left( \frac{x}{b} \right)
\end{equation}
where
\begin{eqnarray}
\label{TDSE-2}
&& E_n = \left( n + \frac{1}{2} \right) \omega(0)     \hspace{1.0cm}  \tau (t) = \int_0^t \frac{d s}{b^2 (s)}       \\   \nonumber
&&\phi_n (x) = \frac{1}{\sqrt{2^n n!}} \left( \frac{ \omega (0)} {\pi b^2} \right)^{1/4} H_n \left(\sqrt{\omega (0)} x \right) e^{-\frac{\omega (0)}{2} x^2 }.
\end{eqnarray}
In Eq. (\ref{TDSE-2}) $H_n (z)$ is $n^{th}$-order Hermite polynomial and $b(t)$ satisfies the Ermakov equation
\begin{equation}
\label{ermakov-1}
\ddot{b} + \omega^2 (t) b = \frac{\omega^2 (0)}{b^3}
\end{equation}
with $b(0) = 1$ and $\dot{b} (0) = 0$. Solution of the Ermakov equation was discussed in Ref. \cite{pinney50}. If $\omega(t)$ is time-independent, $b(t)$ is simply one. 
If $\omega (t)$ is instantly changed as
\begin{eqnarray}
\label{instant-1}
\omega (t) = \left\{                \begin{array}{cc}
                                               \omega_i  & \hspace{1.0cm}  t = 0   \\
                                               \omega_f  & \hspace{1.0cm}  t > 0,
                                               \end{array}            \right.
\end{eqnarray}
then $b(t)$ becomes
\begin{equation}
\label{scale-1}
b(t) = \sqrt{ \frac{\omega_f^2 - \omega_i^2}{2 \omega_f^2} \cos (2 \omega_f t) +  \frac{\omega_f^2 + \omega_i^2}{2 \omega_f^2}}.
\end{equation}
For more general time-dependent case the Ermakov equation should be solved numerically.
Recently, the solution (\ref{scale-1}) is extensively used in Ref. \cite{ghosh17} to discuss the dynamics of entanglement for the sudden quenched states of two site Bose-Hubbard model.
Since TDSE is a linear differential equation, the general solution of TDSE is $\Psi (x, t) = \sum_{n=0}^{\infty} c_n \psi_n (x, t)$ with $\sum_{n=0}^{\infty} |c_n|^2 = 1$. The coefficient $c_n$ is 
determined by making use of the initial conditions.

Using Eqs. (\ref{diag-hamil}) and (\ref{TDSE-1}) the general solution for  TDSE of the  three coupled harmonic oscillators is 
$\Psi (x_1, x_2, x_3 : t) = \sum_{n_1} \sum_{n_+} \sum_{n_-}  c_{n_1,n_+,n_-} \psi_{n_1,n_+,n_-} (x_1, x_2, x_3 : t)$, where
$\sum_{n_1} \sum_{n_+} \sum_{n_-} |c_{n_1,n_+,n_-}|^2 = 1$. In terms of $y_j$ given in Eq. (\ref{change-1}) $\psi_{n_1,n_+,n_-} (x_1, x_2, x_3 : t)$ is expressed as 
\begin{eqnarray}
\label{solu-1}
&& \psi_{n_1,n_+,n_-} (x_1, x_2, x_3 : t) = \frac{1}{\sqrt{2^{n_1+n_++n_-} n_1! n_+! n_-!}} \left( \frac{\omega'_1 \omega'_+ \omega'_- } {\pi^3} \right)^{1/4}                                                                      \\    \nonumber 
&&\hspace{1.0cm} \times e^{-i [E_{n,1} \tau_1 (t) + E_{n,+} \tau_+ (t) + E_{n,-} \tau_- (t)]} 
    e^{\frac{i}{2} \left[ \left( \frac{\dot{b_1}}{b_1} \right) y_1^2 + \left( \frac{\dot{b_+}}{b_+} \right) y_+^2 + \left( \frac{\dot{b_-}}{b_-} \right) y_-^2 \right]}                                                                             \\   \nonumber
&&\hspace{1.0cm} \times H_{n_1} \left( \sqrt{\omega'_1} y_1 \right)  H_{n_+} \left( \sqrt{\omega'_+} y_+ \right)  H_{n_-} \left( \sqrt{\omega'_-} y_- \right) e^{-\frac{1}{2} \left[ \omega'_1 y_1^2 + \omega'_+ y_+^2 + \omega'_-y_-^2 \right]}
\end{eqnarray}
where
\begin{equation}
\label{boso-1}
\omega'_j (t) = \frac{\omega_j (0)}{ b_j^2} \hspace{1.0cm} E_{n,j} = \left(n_j + \frac{1}{2} \right) \omega_j (0) 
\hspace{1.0cm} \tau_j(t) = \int_0^t \frac{d s}{b_j^2 (s)}  
\end{equation}
with $j = 1, \pm$. The scale factors $b_j (t)$ satisfy their own Ermakov equations;
\begin{equation}
\label{ermakov-2}
\ddot{b_j} + \omega_j^2 (t) b_j = \frac{\omega_j^2 (0)}{b_j^3} \hspace{.5cm} (j = 1, \pm)
\end{equation}
with $b_j(0) = 1$ and $\dot{b_j} (0) = 0$. 

In this paper we consider only the vacuum solution $\Psi_0 (x_1, x_2, x_3 : t) = \psi_{0,0,0} (x_1, x_2, x_3 :t)$. Then the density matrix of the whole system 
is given by 
\begin{equation}
\label{density-1}
\rho_{ABC} (x_j : x'_j : t) \equiv \Psi (x_j : t) \Psi^* (x'_j : t) 
= \left( \frac{\omega'_1 \omega'_+ \omega'_-}{\pi^3} \right)^{1/2} \mbox{exp} \left[ -\sum_{i,j = 1}^3 \left( x_i G_{ij} x_j + x'_i G_{ij}^* x'_j \right) \right]
\end{equation}
where $G_{ij} = G_{ji}$ with 
\begin{eqnarray}
\label{density-2}
&&G_{11} = \frac{1}{2} \left[ \frac{v_1}{3} + v_+ A_+^2 (-J_{12} + J_{23} - z)^2 + v_- A_-^2 (-J_{12} + J_{23} + z)^2 \right]     \nonumber   \\
&&G_{22} = \frac{1}{2} \left[ \frac{v_1}{3} + v_+ A_+^2 (J_{12} - J_{13} + z)^2 + v_- A_-^2 (J_{12} - J_{13} - z)^2 \right]     \nonumber       \\
&&G_{33} =  \frac{1}{2} \left[ \frac{v_1}{3} + \left(v_+ A_+^2 + v_- A_-^2 \right) (J_{13} - J_{23})^2 \right] \\   \nonumber
&&G_{12} =  \frac{1}{2} \left[ \frac{v_1}{3} + v_+ A_+^2  (-J_{12} + J_{23} - z) (J_{12} - J_{13} + z) + v_- A_-^2  (-J_{12} + J_{23} + z) (J_{12} - J_{13} - z) \right]                                                                                                                                                      \\   \nonumber   
&&G_{13} =    \frac{1}{2} \left[ \frac{v_1}{3} + \left\{ v_+ A_+^2  (-J_{12} + J_{23} - z) + v_- A_-^2  (-J_{12} + J_{23} + z)  \right\} (J_{13} - J_{23}) \right]                                                                                                                                                        \\   \nonumber  
&&G_{23} =    \frac{1}{2} \left[ \frac{v_1}{3} + \left\{ v_+ A_+^2  (J_{12} - J_{13} + z) + v_- A_-^2  (J_{12} - J_{13} - z)  \right\} (J_{13} - J_{23}) \right].               
\end{eqnarray}
In Eq. (\ref{density-2}) $v_j \hspace{.2cm} (j=1,\pm)$ is defined by 
\begin{equation}
\label{density-3}
v_j = \omega'_j - i \frac{\dot{b_j}}{b_j}.
\end{equation}
In next two sections we discuss on the mixedness and entanglement of the reduced states $\rho_C^{(red)}$ and  $\rho_{BC}^{(red)}$, respectively.

\section{Dynamics of Entanglement between $AB$ and $C$ oscillators}
In this section we assume $AB$ oscillators are inaccessible. Then, the effective state for $C$ oscillator is reduced state, which is given by 
\begin{equation}
\label{reduceC-1}
\rho_C^{(red)} (x_3, x'_3 : t) = \mbox{tr}_{AB} \rho_{ABC} \equiv \int dx_1 dx_2 \rho_{ABC} (x_1, x_2, x_3 : x_1, x_2, x'_3 : t).
\end{equation}
Performing the integration explicitly one can show directly
\begin{equation}
\label{reduceC-2}
\rho_C^{(red)} (x, x' : t) = \left( \frac{\omega'_1 \omega'_+ \omega'_-}{\pi \Omega} \right)^{1/2} 
\mbox{exp} \left[ - \frac{1}{\Omega} \left\{ (R_1 - i I_1) x^2 + (R_1 + i I_1) x'^2 - 2 Y x x' \right\} \right]
\end{equation}
where 
\begin{eqnarray}
\label{reduceC-3}
&&\Omega = \frac{1}{3} \left[ A_+^2 Z_+^2 \omega'_1 \omega'_+ + A_-^2 Z_-^2 \omega'_1 \omega'_- + \omega'_+ \omega'_- \right]  
                                                                                                                                                                                        \nonumber   \\
&&Y = \frac{|v_1|^2}{36} \left( A_+^2 Z_+^2\omega'_+ + A_-^2 Z_-^2 \omega'_- \right) + \frac{(J_{13} - J_{23})^2 \omega'_1}{12}
         \left( A_+^4 Z_+^2 |v_+|^2 + A_-^4 Z_-^2 |v_-|^2 \right)                                                                                              \nonumber    \\
&&\hspace{2.0cm}+ z^2 A_+^2 A_-^2 (J_{13} - J_{23})^4 \left( A_+^2 |v_+|^2 \omega'_- + A_-^2 \omega'_+ |v_-|^2 \right)    \\  \nonumber
&&\hspace{1.0cm}+ \frac{A_+^2 A_-^2}{6} (J_{13} - J_{23})^2 \bigg[ \frac{1}{2} Z_+ Z_- \omega'_1 (v_+ v_-^* + v_+^* v_- ) - z Z_+ \omega'_+ (v_1 v_-^* + v_1^* v_-)                                                    
                                                                                                                                                                                       \\   \nonumber
&& \hspace{6.0cm}+ z Z_- \omega'_- (v_1 v_+^* + v_1^* v_+ )           \bigg]                                                                       \\   \nonumber
&&R_1 = \frac{1}{2} \omega'_1 \omega'_+ \omega'_- + Y                                                                                                 \\   \nonumber
&&I_1 = A_+^2 A_-^2 (J_{13} - J_{23})^2 z \left[ Z_+ \omega'_1 \omega'_+ \frac{\dot{b_-}}{b_-} - Z_- \omega'_1 \frac{\dot{b_+}}{b_+} \omega'_-
+ 2 z \frac{\dot{b_1}}{b_1} \omega'_+ \omega'_-  \right]
\end{eqnarray}
with $Z_{\pm} = 2 J_{12} - J_{13} - J_{23} \pm 2 z$. It is useful to note 
\begin{equation}
\label{useful-2}
Z_+ Z_- = -3  (J_{13} - J_{23})^2.
\end{equation}
It is easy to show 
\begin{equation}
\label{unitarity-1}
\mbox{tr} \left[ \rho_C^{(red)} \right] \equiv \int dx \rho_C^{(red)} (x, x : t) = 1. 
\end{equation}
This guarantees the probability conservation of the $C$-oscillator reduced system. Since $\rho_C^{(red)}$ is a reduced state, it is in general mixed state.
The mixedness of  $\rho_C^{(red)}$ can be measured by 
\begin{equation}
\label{mixedC-1}
\mbox{tr} \left[ \left(\rho_C^{(red)} \right)^2 \right] \equiv \int dx dx' \rho_C^{(red)} (x, x' : t) \rho_C^{(red)} (x', x : t) 
= \sqrt{ \frac{\omega'_1 \omega'_+ \omega'_-}{2 (R_1 + Y)}}.
\end{equation}
Thus, if $Y=0$, $\rho_C^{(red)}$ becomes pure state. It is completely mixed state when $\omega'_1 \omega'_+ \omega'_- = 0$.

The entanglement of $\rho_C^{(red)}$ can be computed by solving the eigenvalue equation
\begin{equation}
\label{eigenC-1}
\int dx' \rho_C^{(red)} (x, x' : t) f_n (x', t) = p_n (t) f_n (x, t).
\end{equation}
One can show that the normalized eigenfunction is 
\begin{equation}
\label{eigenC-2}
f_n (x, t) = \frac{1}{\sqrt{2^n n!}} \left( \frac{\epsilon}{\pi} \right)^{1/4} H_n (\sqrt{\epsilon} x) e^{-\frac{\epsilon}{2} x^2 + i \frac{I_1}{\Omega} x^2}
\end{equation}
where
\begin{equation}
\label{eigenC-3}
\epsilon = 2 \sqrt{\frac{R_1^2 - Y^2}{\Omega^2}},
\end{equation}
and the corresponding eigenvalue is 
\begin{equation}
\label{eigenC-4}
p_n (t) = \left[1 - \xi (t) \right] \xi^n (t)
\end{equation}
where
\begin{equation}
\label{eigenC-5}
\xi (t) = \frac{Y}{R_1 + \sqrt{R_1^2 - Y^2}}.
\end{equation}
 Thus R\'{e}nyi and von Neumann entropies are given by 
 \begin{eqnarray}
 \label{eigenC-6}
 &&S_{\alpha}^C \equiv \frac{1}{1 - \alpha} \ln \mbox{tr} \left[ \left(\rho_C^{(red)} \right)^{\alpha} \right] = \frac{1}{1 - \alpha} 
 \ln \frac{(1 - \xi)^{\alpha}}{1 - \xi^{\alpha}}                                                                                     \\    \nonumber
 &&S_{von}^C = \lim_{\alpha \rightarrow 1} S_{\alpha}^C = -\ln (1 - \xi) - \frac{\xi}{1 - \xi} \ln \xi.
 \end{eqnarray}
These quantities measure the entanglement between $AB$-oscillators and $C$-oscillator. The numerical analysis of these quantities will be 
explored later in the quenched models.

\section{Dynamics of Entanglement between $A$ and $BC$ oscillators}
In this section we assume only $A$ oscillator is inaccessible. Then, the effective state for $BC$ oscillator is reduced state, which is given by 
\begin{equation}
\label{reduceBC-1}
\rho_{BC}^{(red)} (x_2, x_3 : x'_2, x'_3 : t) = \mbox{tr}_A \rho_{ABC} \equiv \int d x_1 \rho_{ABC} (x_1, x_2, x_3 : x_1, x'_2, x'_3 : t).
\end{equation}
After long and tedious calculation one can show 
\begin{equation}
\label{reduceBC-2}
\rho_{BC}^{(red)} (x_1, x_2 : y_1, y_2 : t) = \left( \frac{\omega'_1 \omega'_+ \omega'_-}{\pi^2 A} \right)^{1/2} e^{- \frac{\Gamma}{A}}
\end{equation}
where
\begin{eqnarray}
\label{reduceBC-3}
&& A = G_{11} + G_{11}^* = \frac{\omega'_1}{3} + \omega'_+ A_+^2 (-J_{12} + J_{23} - z)^2 + \omega'_- A_-^2  (-J_{12} + J_{23} + z)^2 
                                                                                                                                                                      \\    \nonumber
&&\Gamma = (\alpha_1 - i \beta_1) x_1^2 +  (\alpha_1 + i \beta_1) y_1^2 +  (\alpha_2 - i \beta_2) x_2^2 + (\alpha_2 + i \beta_2) y_2^2
                                                                                                                                                                       \\    \nonumber
&&\hspace{1.0cm} + 2 (\alpha_3 - i \beta_3) x_1 x_2 + 2  (\alpha_3 + i \beta_3) y_1 y_2 -2 \gamma_{11} x_1 y_1 - 2 \gamma_{22} x_2 y_2
                                                                                                                                                                        \\   \nonumber
&&\hspace{1.0cm} - 2  (\alpha_4 - i \beta_4) x_1 y_2 - 2  (\alpha_4 + i \beta_4) x_2 y_1.
\end{eqnarray}
In $\Gamma$ $\alpha_i$, $\beta_i$, and $\gamma_{ij}$ are all real quantities and have long expressions. Their explicit expressions are given in appendix A.
Here, we present several useful formula
\begin{eqnarray}
\label{useful-3}
&&\alpha_1 - \gamma_{11} = \frac{A_+^2}{6} Z_+^2 \omega'_1 \omega'_+ + \frac{A_-^2}{6} Z_-^2 \omega'_1 \omega'_- 
+ 2 A_+^2 A_-^2 z^2 (J_{13} - J_{23})^2 \omega'_+ \omega'_-                                                     \\    \nonumber
&&\alpha_2 - \gamma_{22} = \frac{A_+^2}{6} Y_+^2 \omega'_1 \omega'_+ + \frac{A_-^2}{6} Y_-^2 \omega'_1 \omega'_- 
+ 2 A_+^2 A_-^2 z^2 (J_{13} - J_{23})^2 \omega'_+ \omega'_-                                                      \\    \nonumber
&&\alpha_3 - \alpha_4 = \frac{A_+^2}{6} Y_+ Z_+ \omega'_1 \omega'_+ + \frac{A_-^2}{6} Y_- Z_- \omega'_1 \omega'_- 
- 2 A_+^2 A_-^2 z^2 (J_{13} - J_{23})^2 \omega'_+ \omega'_-
\end{eqnarray}
where $Y_{\pm} = J_{12} + J_{13} - 2 J_{23} \pm z$. Using Eq. (\ref{useful-3}) it is straight to show 
\begin{equation}
\label{useful-4}
(\alpha_1 - \gamma_{11}) (\alpha_2 - \gamma_{22}) - (\alpha_3 - \alpha_4)^2 = \frac{\omega'_1 \omega'_+ \omega'_- A}{4}.
\end{equation}
Then, it is easy to show
\begin{equation}
\label{unitarity-2}
\mbox{tr} \left[ \rho_{BC}^{(red)} \right] \equiv \int dx_1 dx_2 \rho_{BC}^{(red)} (x_1, x_2 : x_1, x_2 : t) = 1.
\end{equation}
Also one can compute the measure of the mixedness for $\rho_{BC}^{(red)}$, which is 
\begin{eqnarray}
\label{mixedBC-1}
&&\mbox{tr} \left[ \left( \rho_{BC}^{(red)} \right)^2 \right] 
\equiv \int dx_1 dx_2 dy_1 dy_2 \rho_{BC}^{(red)} (x_1, x_2 : y_1, y_2 : t) \rho_{BC}^{(red)} (y_1, y_2, x_1, x_2 : t)       \nonumber    \\
&& \hspace{5.0cm}= \frac{\omega'_1 \omega'_+ \omega'_- A}{4} \sqrt{ \frac{\alpha_2^2 - \gamma_{22}^2}{n_1^2 - n_2^2}}
\end{eqnarray}
where
\begin{eqnarray}
\label{mixedBC-2}
&&n_1 = \alpha_1 (\alpha_2^2 - \gamma_{22}^2) - \alpha_2 (\alpha_3^2 + \alpha_4^2) + 2 \gamma_{22} \alpha_3 \alpha_4     \\     \nonumber
&&n_2 = \gamma_{11}  (\alpha_2^2 - \gamma_{22}^2) + \gamma_{22}  (\alpha_3^2 + \alpha_4^2)  -2 \alpha_2 \alpha_3 \alpha_4.
\end{eqnarray}

In order to discuss the entanglement between $A$-oscillator and $BC$-oscillator we should solve the eigenvalue equation
\begin{equation}
\label{eigenBC-1}
\int dy_1 dy_2  \rho_{BC}^{(red)} (x_1, x_2 : y_1, y_2 : t) f_{mn} (y_1, y_2 : t) = p_{mn} (t) f_{mn} (x_1, x_2 : t).
\end{equation}

If the oscillator $A$ is accessible, one can compute the  R\'{e}nyi and von Neumann entropies of  $\rho_{BC}^{(red)}$ more easily without solving 
Eq. (\ref{eigenBC-1})  because the total state $\rho_{ABC}$ is pure. From Schmidt decomposition we know that the eigenvalue spectrum and hence, entropies of $\rho_{BC}^{(red)}$ are exactly the same with those of $\rho_{A}^{(red)}$. Since, however, the oscillator $A$ is assumed to be inaccessible, 
we should compute the entropies of  $\rho_{BC}^{(red)}$ by solving Eq. (\ref{eigenBC-1}) directly. For completeness we compute the  R\'{e}nyi and von Neumann entropies of  $\rho_{BC}^{(red)}$ again in appendix B by making use of  $\rho_{A}^{(red)}$.

In order to solve the eigenvalue equation (\ref{eigenBC-1}) we define 
\begin{equation}
\label{eugenBC-2}
f_{mn} (x_1, x_2 : t) = e^{\frac{i}{A} \left(\beta_1 x_1^2 + \beta_2 x_2^2 + 2 \beta_3 x_1 x_2 \right)} g_{mn} (x_1, x_2 : t).
\end{equation}
Then, Eq. (\ref{eigenBC-1}) reduces to 
\begin{eqnarray}
\label{eigenBC-3}
&& C_{{\cal N}}  e^{-\frac{1}{A} \left(\alpha_1 x_1^2 + \alpha_2 x_2^2 + 2 \alpha_3 x_1 x_2 \right)}                                                                                                  
                                                                                                                                                         \\    \nonumber
&&  \times\int dy_1 dy_2  e^{-\frac{1}{A} \left(\alpha_1 y_1^2 + \alpha_2 y_2^2 + 2 \alpha_3 y_1 y_2 - 2 a y_1 - 2 b y_2\right)} g_{mn} (y_1, y_2 : t)
                                                                        = p_{mn} (t) g_{mn} (x_1, x_2 : t)
\end{eqnarray}
where 
\begin{equation}
\label{eigenBC-4}
a (t) = \gamma_{11} x_1 + (\alpha_4 + i \beta_4) x_2              \hspace{1.0cm}          
b (t) = (\alpha_4 - i \beta_4) x_1 + \gamma_{22} x_2
\end{equation}
and $C_{{\cal N}}$ is a multiplicative constant. From now on the multiplicative constant will be absorbed into  $C_{{\cal N}}$ although it is changed due to 
Jacobian factors. It can be fixed after calculation is complete by making use of Eq. (\ref{unitarity-2}).

Now, we define new coordinates
\begin{eqnarray}
\label{eigenBC-5}
&&\tilde{y_1} = \frac{1}{{\cal N}} \left[ 2 \alpha_3 y_1 + \left\{ \eta - (\alpha_1 - \alpha_2) \right\} y_2 \right],     \hspace{0.1cm}
\tilde{y_2} = \frac{1}{{\cal N}} \left[ - \left\{ \eta - (\alpha_1 - \alpha_2) \right\} y_1 + 2 \alpha_3 y_2 \right]        \\   \nonumber
&&\tilde{x_1} = \frac{1}{{\cal N}} \left[ 2 \alpha_3 x_1 + \left\{ \eta - (\alpha_1 - \alpha_2) \right\} x_2 \right],     \hspace{0.1cm}
\tilde{x_2} = \frac{1}{{\cal N}} \left[ - \left\{ \eta - (\alpha_1 - \alpha_2) \right\} x_1 + 2 \alpha_3 x_2 \right]           
\end{eqnarray}
where
\begin{equation}
\label{eigenBC-6}
\eta = \sqrt{ (\alpha_1 - \alpha_2)^2 + 4 \alpha_3^2}     \hspace{1.0cm} {\cal N}^2 = 2 \eta [ \eta - (\alpha_1 - \alpha_2)].
\end{equation}
Then the eigenvalue equation (\ref{eigenBC-3}) becomes
\begin{eqnarray}
\label{eigenBC-7}
&& C_{{\cal N}} e^{-\frac{1}{A} \left( \eta_+ \tilde{x_1}^2 + \eta_- \tilde{x_2}^2 \right)}
                                                                                                                                                                       \\    \nonumber
&&\times \int d \tilde{y_1} d \tilde{y_2}  e^{-\frac{1}{A} \left( \eta_+ \tilde{y_1}^2 + \eta_- \tilde{y_2}^2 -2 \sum_{i,j=1}^2 c_{ij} \tilde{x_i} \tilde{y_j} \right)}
g_{mn} (\tilde{y_1}, \tilde{y_2} : t) = p_{mn} (t) g_{mn} (\tilde{x_1}, \tilde{x_2} : t)
\end{eqnarray}
where 
\begin{equation}
\label{eigenBC-8}
\eta_{\pm} = \frac{(\alpha_1 + \alpha_2) \pm \eta}{2}
\end{equation}
and 
\begin{eqnarray}
\label{eigenBC-9}
&&c_{11} = \frac{1}{{\cal N}^2} \left[ 4 \alpha_3^2 \gamma_{11} + 4 \alpha_3 \alpha_4 \left\{ \eta - (\alpha_1 - \alpha_2) \right\} 
+ \gamma_{22}  \left\{ \eta - (\alpha_1 - \alpha_2) \right\}^2 \right]                                   \\     \nonumber
&&c_{22} = \frac{1}{{\cal N}^2} \left[ 4 \alpha_3^2 \gamma_{22} - 4 \alpha_3 \alpha_4 \left\{ \eta - (\alpha_1 - \alpha_2) \right\} 
+ \gamma_{11}  \left\{ \eta - (\alpha_1 - \alpha_2) \right\}^2 \right]                                    \\    \nonumber
&&c_{12} = \frac{1}{{\cal N}^2} \left[ 4 \alpha_3^2 \alpha_4 - 2 \alpha_3 (\gamma_{11} - \gamma_{22}) \left\{ \eta - (\alpha_1 - \alpha_2)
\right\} - \alpha_4 \left\{ \eta - (\alpha_1 - \alpha_2) \right\}^2  - i \beta_4 {\cal N}^2 \right]     
\end{eqnarray}
with $c_{21} = c_{12}^*$. In order to simplify Eq. (\ref{eigenBC-7}) some more we define new coordinates again as 
\begin{eqnarray}
\label{eigenBC-10}
&& \bar{x_1} = \sqrt{\eta_+} \tilde{x_1}     \hspace{1.0cm}  \bar{x_2} = \sqrt{\eta_-} \tilde{x_2}                       \\    \nonumber
&& \bar{y_1} = \sqrt{\eta_+} \tilde{y_1}     \hspace{1.0cm}  \bar{y_2} = \sqrt{\eta_-} \tilde{y_2}.
\end{eqnarray}
Then Eq. (\ref{eigenBC-7}) becomes 
\begin{eqnarray}
\label{eigenBC-11}
&& C_{{\cal N}} e^{-\frac{1}{A} \left(\bar{x_1}^2 +  \bar{x_2}^2 \right)}
 \int d \bar{y_1} d \bar{y_2}  e^{-\frac{1}{A} \left( \bar{y_1}^2 +  \bar{y_2}^2 -2 \sum_{i,j=1}^2 \kappa_{ij} \bar{x_i} \bar{y_j} \right)}
g_{mn} (\bar{y_1}, \bar{y_2} : t)                                           \nonumber     \\
&&\hspace{5.0cm}  = p_{mn} (t) g_{mn} (\bar{x_1}, \bar{x_2} : t)
\end{eqnarray}
where
\begin{equation}
\label{eigenBC-12}
\kappa_{11} = \frac{c_{11}}{\eta_+}   \hspace{1.0cm}  \kappa_{22} = \frac{c_{22}}{\eta_-}   \hspace{1.0cm}
\kappa_{12} = \frac{c_{12}}{\sqrt{\eta_+ \eta_-}}     \hspace{1.0cm}  \kappa_{21} = \frac{c_{21}}{\sqrt{\eta_+ \eta_-}}.
\end{equation}
Since $\kappa_{ij}$ is a hermitian matrix, it can be diagonalized by introducing an appropriate unitary matrix. Using the unitary matrix we define new
coordinates finally as 
\begin{eqnarray}
\label{eigenBC-13}
&&X_1 = \frac{1}{{\cal N}_{\kappa}} \left[ 2 \kappa_{21} \bar{x_1} + \left\{\chi - (\kappa_{11} - \kappa_{22}) \right\} \bar{x_2} \right]
                                                                                                                                                      \\    \nonumber
&&X_2 = \frac{1}{{\cal N}_{\kappa}} \left[-\left\{\chi - (\kappa_{11} - \kappa_{22}) \right\} \bar{x_1} + 2 \kappa_{12} \bar{x_2}  \right]
                                                                                                                                                       \\    \nonumber
&&Y_1 = \frac{1}{{\cal N}_{\kappa}} \left[ 2 \kappa_{21} \bar{y_1} + \left\{\chi - (\kappa_{11} - \kappa_{22}) \right\} \bar{y_2} \right]
                                                                                                                                                      \\    \nonumber
&&Y_2 = \frac{1}{{\cal N}_{\kappa}} \left[-\left\{\chi - (\kappa_{11} - \kappa_{22}) \right\} \bar{y_1} + 2 \kappa_{12} \bar{y_2}  \right]
\end{eqnarray}
where
\begin{equation}
\label{eigenBC-14}
\chi = \sqrt{(\kappa_{11} - \kappa_{22})^2 + 4 |\kappa_{12}|^2}     \hspace{1.0cm}
{\cal N}_{\kappa}^2 = 2 \chi \left[ \chi - (\kappa_{11} - \kappa_{22}) \right].
\end{equation}
In terms of the new coordinates Eq. (\ref{eigenBC-11}) is simplified as 
\begin{eqnarray}
\label{eigenBC-15}
&&  C_{{\cal N}} e^{-\frac{1}{A} (X_1^2 + X_2^2)}     \\      \nonumber
&& \times \int dY_1 dY_2  e^{-\frac{1}{A} \left[Y_1^2 + Y_2^2 - 2 (\chi_+ X_1 Y_1 + \chi_- X_2 Y_2) \right]} g_{mn} (Y_1, Y_2 : t)     
= p_{mn} (t)  g_{mn} (X_1, X_2 : t)
\end{eqnarray}
where
\begin{equation}
\label{eigenBC-16}
\chi_{\pm} = \frac{1}{2} \left[ (\kappa_{11} + \kappa_{22}) \pm \chi \right].
\end{equation}
Then Eq. (\ref{eigenBC-15}) is divided into two single variable eigenvalue equations as 
\begin{eqnarray}
\label{eigenBC-17}
&&L_1 e^{-\frac{1}{A} X_1^2} \int d Y_1 e^{-\frac{1}{A} \left( Y_1^2 - 2 \chi_+ X_1 Y_1 \right)} g_{1,m} (Y_1, t) = q_{1,m} (t) g_{1,m}(X_1, t)
                                                                                                                                                                             \\    \nonumber
&&L_2 e^{-\frac{1}{A} X_2^2} \int d Y_2 e^{-\frac{1}{A} \left( Y_2^2 - 2 \chi_- X_2 Y_2 \right)} g_{2,n} (Y_2, t) = q_{2,n} (t) g_{2,n}(X_2, t)
\end{eqnarray}
where
\begin{eqnarray}
\label{eigenBC-18}
&& L_1 L_1 =  C_{{\cal N}} \hspace{1.0cm}
p_{mn} (t) =  q_{1,m} (t) q_{2,n} (t)                                                                                                                          \\     \nonumber 
&&\hspace{1.0cm}   g_{mn} (X_1, X_2 : t) =  g_{1,m}(X_1, t)  g_{2.n}(X_2, t).
\end{eqnarray}
Each eigenvalue equation in Eq. (\ref{eigenBC-17}) can be solved easily. Then, the normalized eigenfunction of $\rho_{BC}^{(red)}$ is 
\begin{eqnarray}
\label{eigenBC-19}
&&g_{mn} (X_1, X_2 : t)                                                  \\     \nonumber
&&= \left[ \frac{1}{\sqrt{2^m m!}} \left( \frac{\epsilon_1}{\pi} \right)^{1/4} H_m (\sqrt{\epsilon_1} X_1) e^{-\frac{\epsilon_1}{2} X_1^2} \right]
\left[ \frac{1}{\sqrt{2^n n!}} \left( \frac{\epsilon_2}{\pi} \right)^{1/4} H_n (\sqrt{\epsilon_2} X_2) e^{-\frac{\epsilon_2}{2} X_2^2} \right]
\end{eqnarray}
and the corresponding eigenvalue is 
\begin{equation}
\label{eigenBC-20}
p_{mn} (t) = \left[L_1 \sqrt{\frac{\pi}{\frac{1}{A} + \frac{\epsilon_1}{2}}} \left( \frac{\frac{1}{A} - \frac{\epsilon_1}{2}} {\frac{1}{A} +  \frac{\epsilon_1}{2}} \right)^{m/2} \right]
\left[L_2 \sqrt{\frac{\pi}{\frac{1}{A} + \frac{\epsilon_2}{2}}} \left( \frac{\frac{1}{A} - \frac{\epsilon_2}{2}} {\frac{1}{A} +  \frac{\epsilon_2}{2}} \right)^{n/2} \right]
\end{equation}
where
\begin{equation}
\label{eigenBC-21}
\frac{\epsilon_1}{2} = \frac{1}{A} \sqrt{1 - \chi_+^2}                \hspace{2.0cm}
\frac{\epsilon_2}{2} = \frac{1}{A} \sqrt{1 - \chi_-^2}.  
\end{equation}
Since Eq. (\ref{unitarity-2}) guarantees $\sum_{mn} p_{m,n} (t) = 1$, one can fix $C_{{\cal N}} = L_1 L_2$. Then, $p_{mn} (t)$ becomes
\begin{equation}
\label{eigenBC-22}
p_{mn} (t) = (1 - \xi_1) \xi_1^m (1 - \xi_2) \xi_2^n
\end{equation}
where
\begin{equation}
\label{eigenBC-23}
\xi_1 = \frac{\chi_+}{1 + \sqrt{1 - \chi_+^2}}     \hspace{2.0cm}  \xi_2 = \frac{\chi_-}{1 + \sqrt{1 - \chi_-^2}}.
\end{equation}
Thus R\'{e}nyi and von Neumann entropies for $\rho_{BC}^{(red)}$ are given by 
\begin{eqnarray}
\label{eigenBC-24}
&&S_{\alpha}^{BC} \equiv \frac{1}{1 - \alpha} \ln \mbox{tr} \left[ \left( \rho_{BC}^{(red)} \right)^{\alpha} \right] = S_{1,\alpha} + S_{2,\alpha}
                                                                                                                                                                             \\     \nonumber
&&S_{von}^{BC} \equiv \lim_{\alpha \rightarrow 1} S_{\alpha}^{BC} = S_{1,von} + S_{2,von}
\end{eqnarray}
where
\begin{eqnarray}
\label{eigenBC-25}
&&S_{1,\alpha} = \frac{1}{1 - \alpha} \ln \frac{(1 - \xi_1)^{\alpha}}{1 - \xi_1^{\alpha}}    \hspace{2.0cm}
S_{2,\alpha} = \frac{1}{1 - \alpha} \ln \frac{(1 - \xi_2)^{\alpha}}{1 - \xi_2^{\alpha}}                             \\    \nonumber
&&S_{1,von} = - \ln (1 - \xi_1) - \frac{\xi_1}{1 - \xi_1} \ln \xi_1                                          \hspace{1.0cm}
S_{2,von} = - \ln (1 - \xi_2) - \frac{\xi_2}{1 - \xi_2} \ln \xi_2. 
\end{eqnarray}

\section{Numerical Analysis : Sudden Quenched Models}

\begin{figure}[ht!]
\begin{center}
\includegraphics[height=5.0cm]{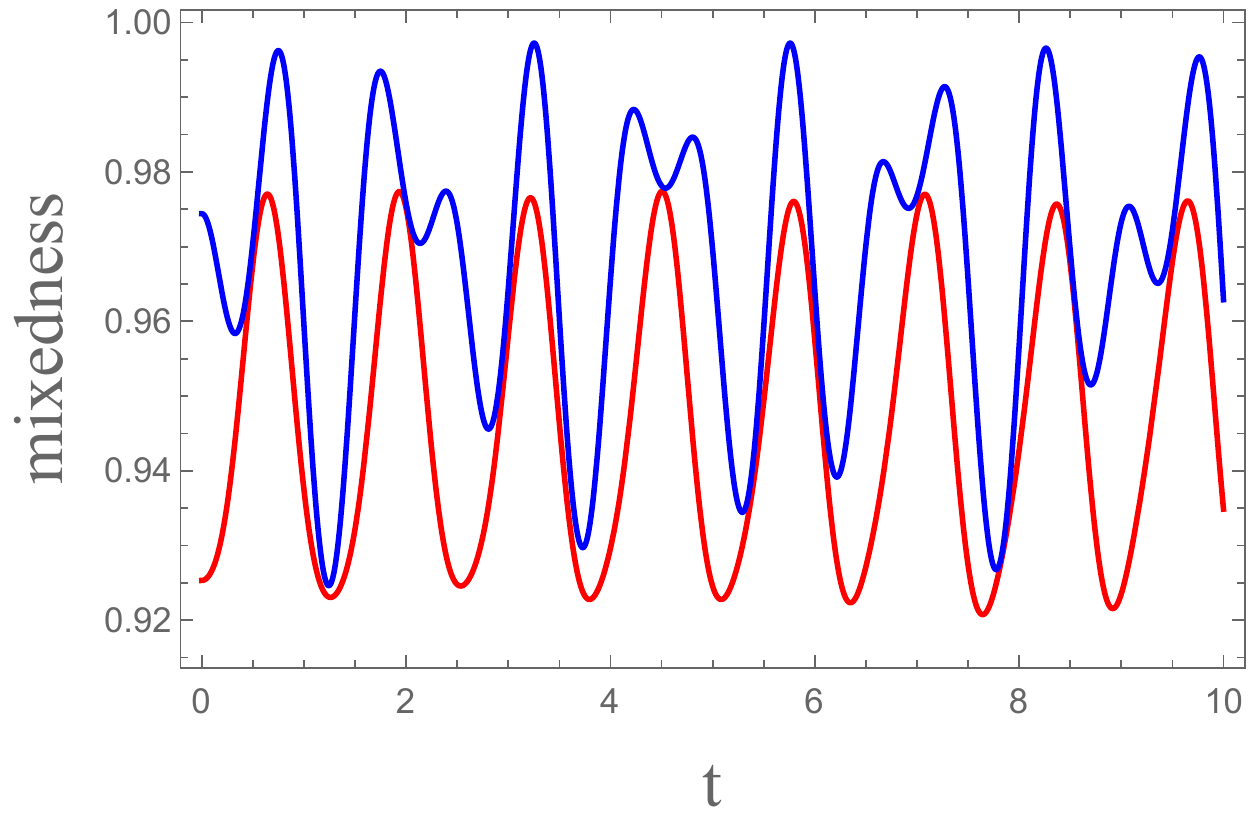} 
\includegraphics[height=5.0cm]{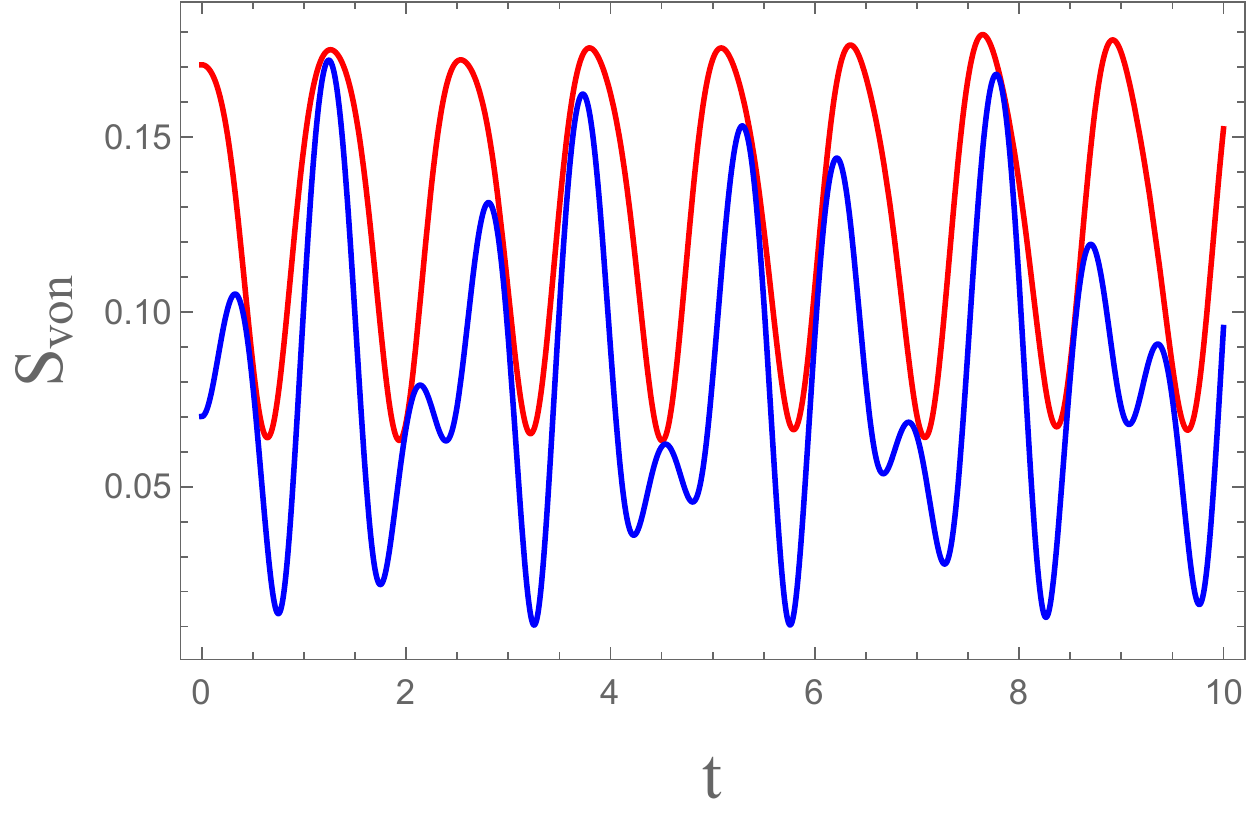}
\includegraphics[height=5.0cm]{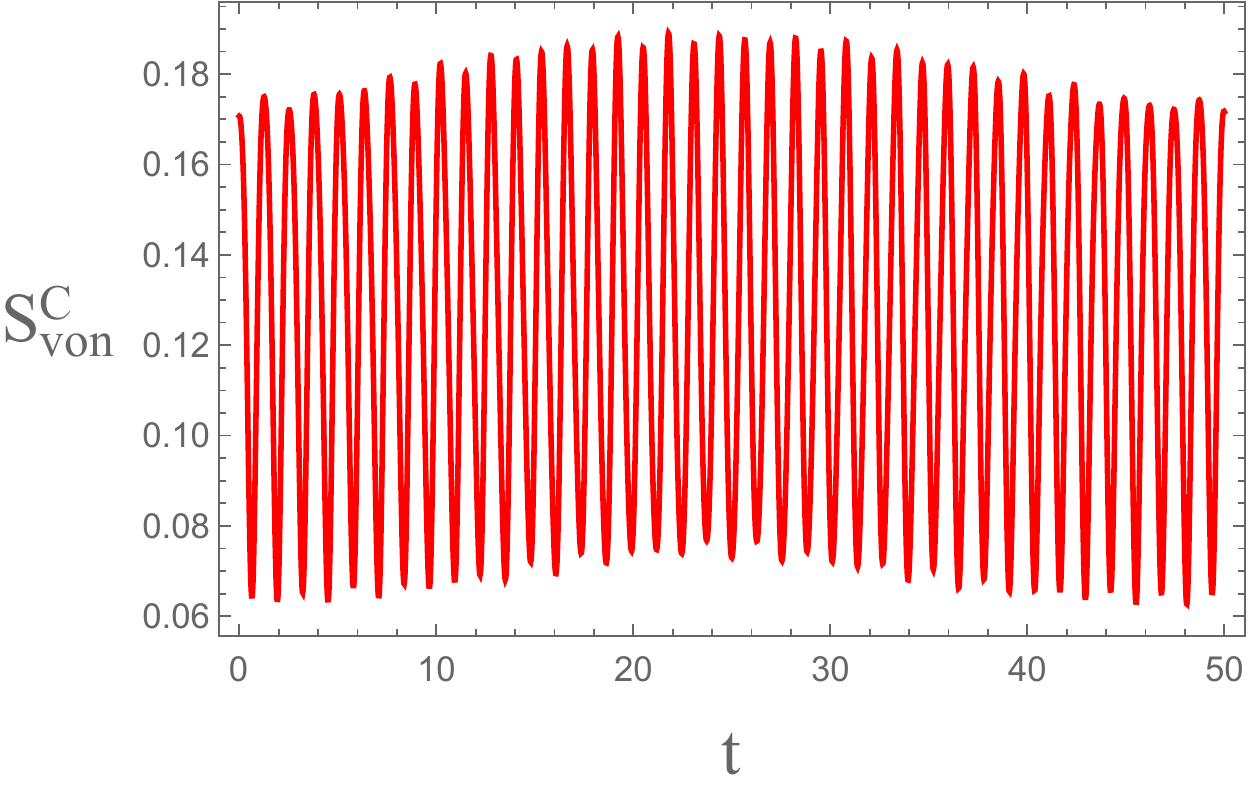}
\includegraphics[height=5.0cm]{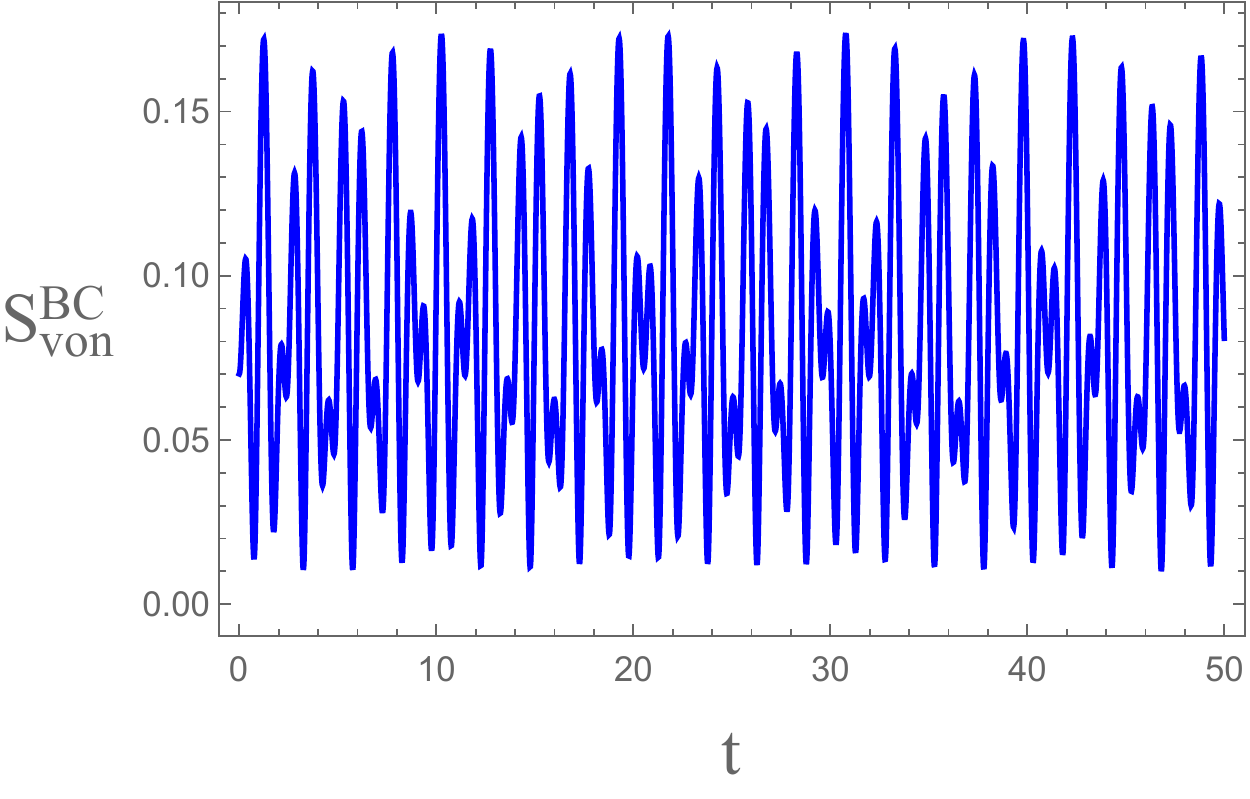}

\caption[fig1]{(Color online) The time-dependence of mixedness (Fig. 1(a)) and von Neumann entropy (Fig. 1(b)) when the quenched parameters are 
chosen as $K_{0,i} = 4$, $K_{0,f} = 6$, $J_{12,i} = 1$, $J_{12,f} = 2$, $J_{13,i} = 3$, $J_{13, f} = 4$, $J_{23, i} = 8$, and $J_{23,f} = 7$. 
The red and blue lines correspond to $\rho_C^{(red)}$ and $\rho_{BC}^{(red)}$ respectively. In order to examine the dependence of multi-frequencies
we plot the time-dependence of  von Neumann entropy for  $\rho_C^{(red)}$ (Fig. 1(c)) and $\rho_{BC}^{(red)}$ (Fig. 1(d)) along the long time interval.
 }
\end{center}
\end{figure}
Using the results of the previous sections we examine in this section the dynamics of the mixedness and entanglement for 
$\rho_C^{(red)}$ and $\rho_{BC}^{(red)}$. Although we can consider more general time-dependent cases by solving the Ermakov equation
(\ref{ermakov-1}) numerically, we confine ourselves in this section into the more simple sudden quenched model, where the time-dependence of 
frequency parameter $K_0 (t)$ and coupling constants $J_{ij} (t)$ arises from abrupt change at $t=0$ such as 
\begin{eqnarray}
\label{quenched-1}
{K}_0 (t) = \left\{                \begin{array}{cc}
                                                K_{0, i}  & \hspace{0.25cm} t = 0   \\
                                                K_{0, f}    & \hspace{0.5cm} t > 0
                                               \end{array}            \right.         \hspace{1.0cm}
{J}_{ij} (t) = \left\{                \begin{array}{cc}
                                               J_{ij, i}  & \hspace{0.25cm} t = 0   \\
                                               J_{ij, f}   & \hspace{0.25cm}  t > 0.
                                               \end{array}            \right.
\end{eqnarray}
Then, $\omega_1 (t)$ and $\omega_{\pm} (t)$ defined in the diagonal Hamiltonian (\ref{diag-hamil}) become
\begin{eqnarray}
\label{quenched-2}
&&\omega_{1,i} = \sqrt{K_{0,i}}    \hspace{2.0cm}   \omega_{1,f} = \sqrt{K_{0,f}}       \\    \nonumber
&&\omega_{\pm, i} = \sqrt{K_{0,i} + J_{12,i} + J_{13, i} + J_{23,i} \pm z_i}          \\    \nonumber
&&\omega_{\pm, f} = \sqrt{K_{0,f} + J_{12,f} + J_{13, f} + J_{23,f} \pm z_f}
\end{eqnarray}
where $z_i$ and $z_f$ are initial and later-time values of $z(t)$. Thus the scale factors $b_{\alpha} (t) \hspace{.2cm} (\alpha = 1, \pm)$ are 
given by 
\begin{equation}
\label{quenched-3}
b_{\alpha} (t) = \sqrt{\frac{\omega_{\alpha, f}^2 - \omega_{\alpha,i}^2}{2 \omega_{\alpha, f}^2} \cos \left( 2 \omega_{\alpha, f} t \right) + 
\frac{\omega_{\alpha, f}^2 + \omega_{\alpha,i}^2}{2 \omega_{\alpha, f}^2}}.
\end{equation}
The trigonometric functions in $b_{\alpha} (t)$ make oscillatory behavior in the dynamics of mixedness and entanglement. 

\begin{figure}[ht!]
\begin{center}
\includegraphics[height=5.0cm]{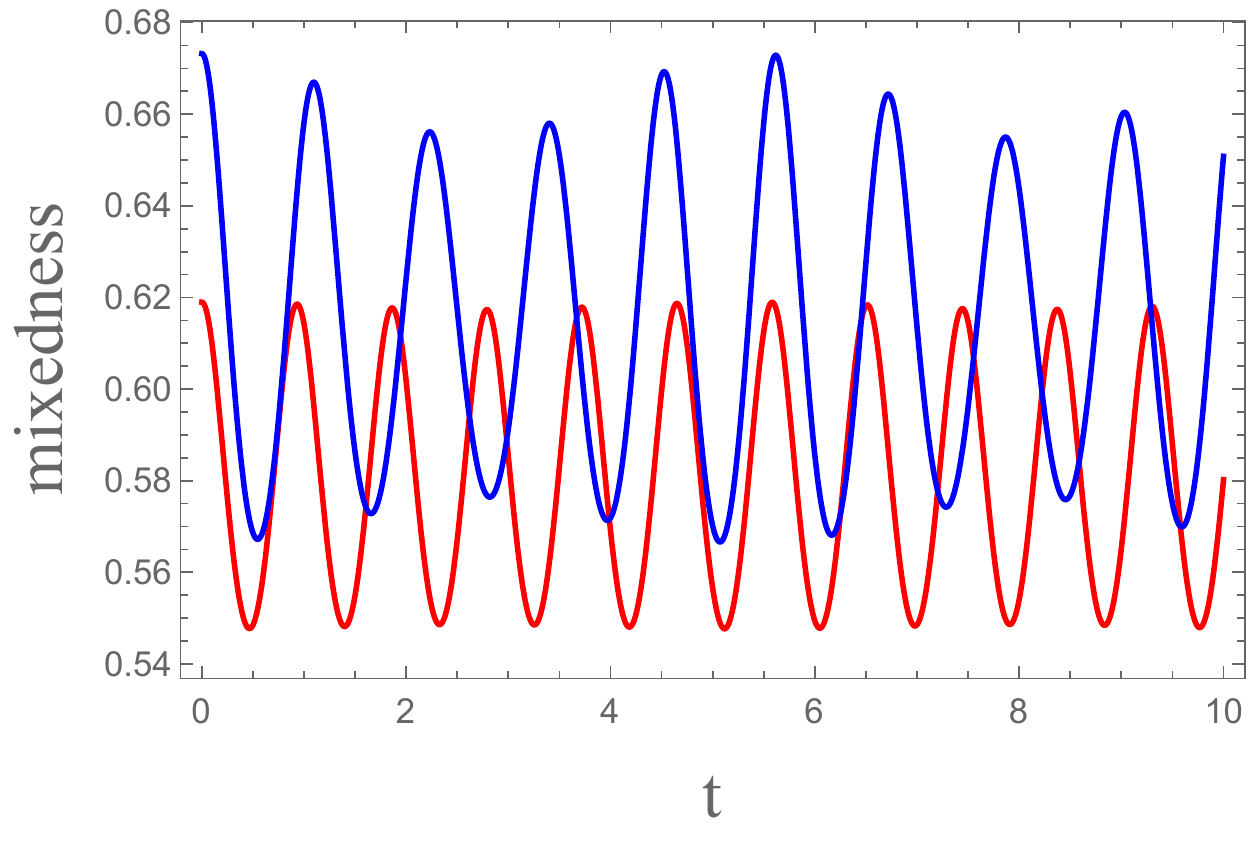} 
\includegraphics[height=5.0cm]{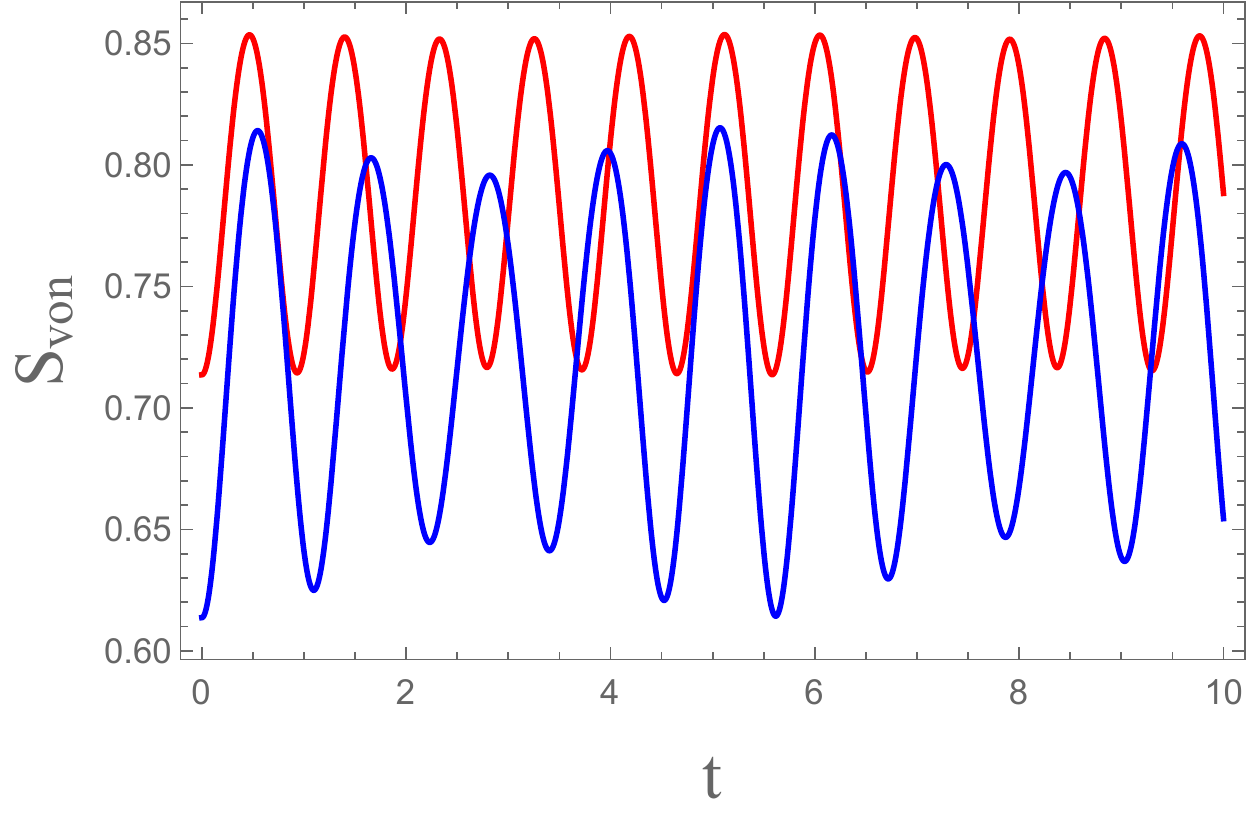}
\includegraphics[height=5.0cm]{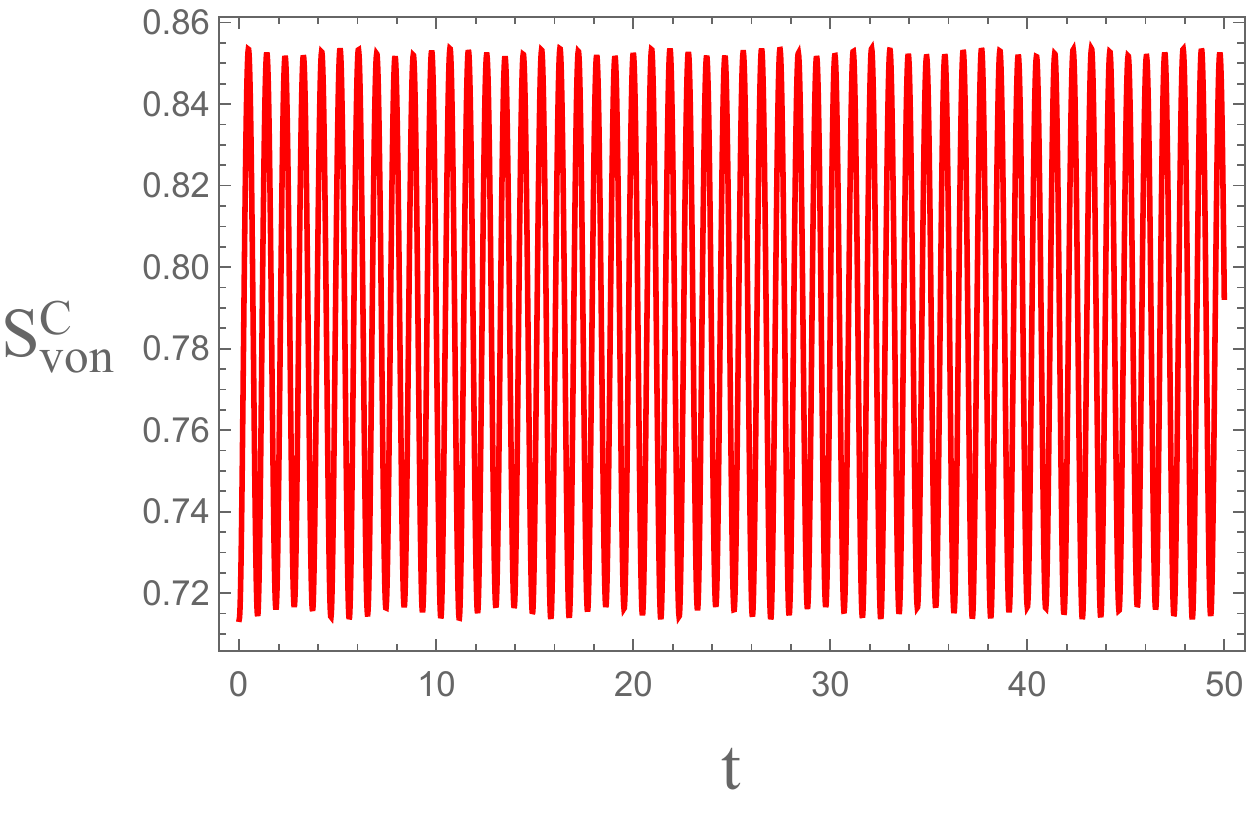}
\includegraphics[height=5.0cm]{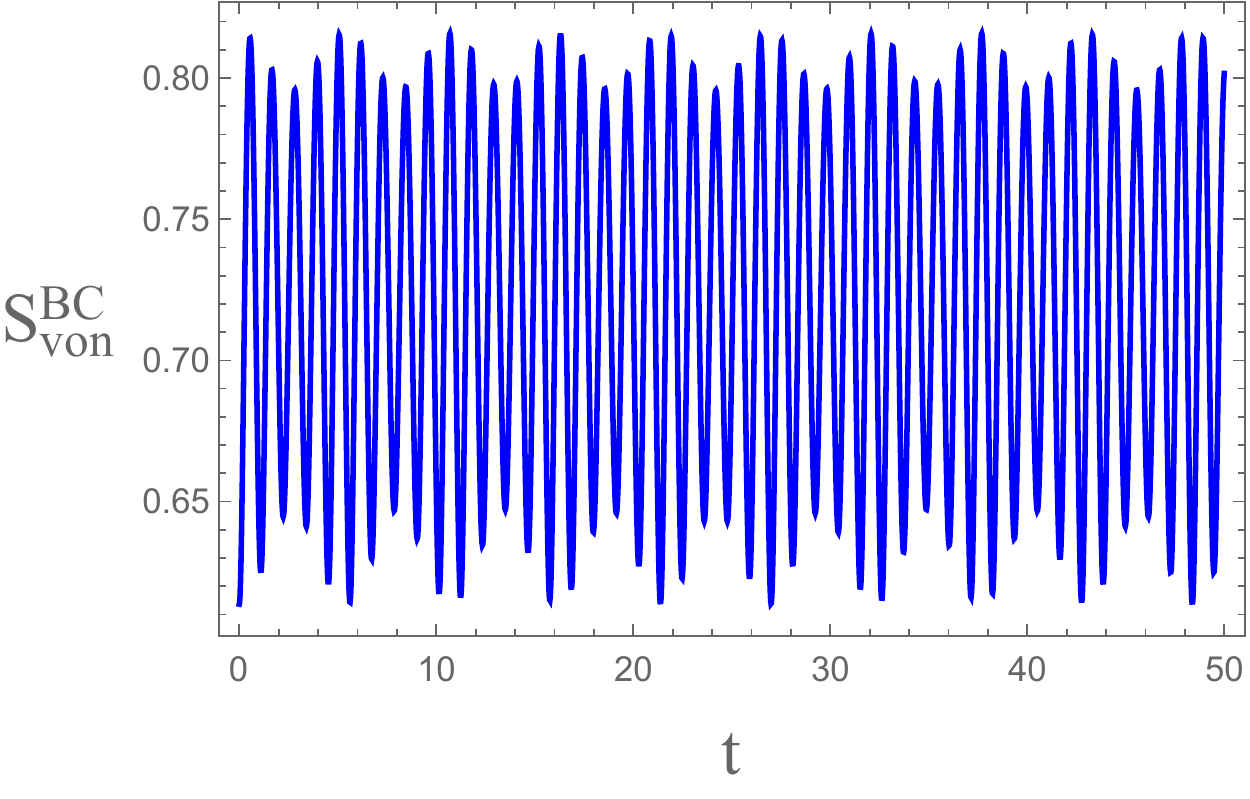}

\caption[fig2]{(Color online) The time-dependence of mixedness (Fig. 2(a)) and von Neumann entropy (Fig. 2(b)) when the quenched parameters are 
chosen as  $K_{0,i} = 0.1$,  $K_{0,f} = 0.1$,  $J_{12,i} = 1$, $J_{12,f} = 2$, $J_{13,i} = 2.5$, $J_{13,f} = 3.5$, $J_{23,i} = 3$, and $J_{23,f} = 4$.
The red and blue lines correspond to $\rho_C^{(red)}$ and $\rho_{BC}^{(red)}$ respectively. In order to examine the dependence of multi-frequencies
we plot the time-dependence of  von Neumann entropy for  $\rho_C^{(red)}$ (Fig. 2(c)) and $\rho_{BC}^{(red)}$ (Fig. 2(d)) along the long time interval. 
Since constant $K_0$ gives $b_1 (t) = 1$,  the effect of multi-frequency seems to be reduced in Fig. 2(c) and Fig. 2(d) compared to Fig. 1(c) and Fig. 1(d).  }
\end{center}
\end{figure}
First, we choose $K_{0,i} = 4$, $K_{0,f} = 6$, $J_{12,i} = 1$, $J_{12,f} = 2$, $J_{13,i} = 3$, $J_{13, f} = 4$, $J_{23, i} = 8$, and $J_{23,f} = 7$. 
In this case $\omega_{1,i} = 2$,  $\omega_{1,f} = 2.45$,  $\omega_{+,i} = 4.72$,  $\omega_{+,f} = 4.83$,  $\omega_{-,i} = 3.12$, and 
$\omega_{-,f} = 3.83$. The time-dependence of $\mbox{tr} \left[ \left(\rho_{BC}^{(red)} \right)^2 \right]$ (blue line) and 
 $\mbox{tr} \left[ \left(\rho_{C}^{(red)} \right)^2 \right]$ (red line) is plotted in Fig. 1(a). As expected both exhibit oscillatory behavior in time. In the 
 full-time range  $\mbox{tr} \left[ \left(\rho_{BC}^{(red)} \right)^2 \right]$ is larger than  $\mbox{tr} \left[ \left(\rho_{C}^{(red)} \right)^2 \right]$.
 This means $\rho_{C}^{(red)}$ is more mixed than $\rho_{BC}^{(red)}$. This can be understood as follows. The total state $\rho_{ABC}$ in 
 Eq. (\ref{density-1}) is pure state. Since $\rho_{C}^{(red)}$ and $\rho_{BC}^{(red)}$ are effective quantum states when two or one oscillator is lost
 respectively, one can expect  $\rho_{C}^{(red)}$ is more mixed than  $\rho_{BC}^{(red)}$. Fig. 1(b) shows the time-dependence of 
 $S_{von}^C$ (red line) and  $S_{von}^{BC}$ (blue line). As expected both exhibit oscillatory behavior in time due to $b_{\alpha} (t)$. In the full-time 
 range  $S_{von}^C$ is larger than  $S_{von}^{BC}$. The multi-frequency dependence of von Neumann and R\'{e}nyi entropies can be seen explicitly if we
 increases the time domain. Fig. 1(c) and Fig. 1(d) are time-dependence of  $S_{von}^C$ and  $S_{von}^{BC}$ in $0 \leq t \leq 50$. These figures clearly
 exhibit the multi-frequency dependence. 

Next, we choose time-independent $K_0$ as $K_0 = 0.1$. Thus, $\omega_1$ is also time-independent as $\omega_1 = 0.316$. The remaining parameters
are chosen as $J_{12,i} = 1$, $J_{12,f} = 2$, $J_{13,i} = 2.5$, $J_{13,f} = 3.5$, $J_{23,i} = 3$, and $J_{23,f} = 4$. In this case $\omega_{\pm}$ become 
$\omega_{+,i} = 2.90$, $\omega_{-,i} = 2.19$, $\omega_{+,f} = 3.38$, and $\omega_{-,f} = 2.79$.  With these parameters the dynamics of mixedness 
and entanglement are plotted in Fig. 2. In Fig. 2(a) the time-dependence of  $\mbox{tr} \left[ \left(\rho_{BC}^{(red)} \right)^2 \right]$ (blue line) and 
$\mbox{tr} \left[ \left(\rho_{C}^{(red)} \right)^2 \right]$ (red line) is plotted. Unlike the previous case  
$\mbox{tr} \left[ \left(\rho_{BC}^{(red)} \right)^2 \right]$ is not always larger than  $\mbox{tr} \left[ \left(\rho_{C}^{(red)} \right)^2 \right]$ in
the full-time range even though the average value of  $\mbox{tr} \left[ \left(\rho_{BC}^{(red)} \right)^2 \right]$ is larger than that of 
$\mbox{tr} \left[ \left(\rho_{C}^{(red)} \right)^2 \right]$. The time-dependence of  $S_{von}^C$ (red line) and  $S_{von}^{BC}$ (blue line) is plotted in 
Fig. 2(b). Similarly,  $S_{von}^C$ is not always larger than  $S_{von}^{BC}$ even though it is right in most time interval. In order to examine the effect 
of constant $\omega_1$ we plot  $S_{von}^C$ (Fig. 2(c)) and  $S_{von}^{BC}$ (Fig. 2(d)) with a long range of time ($0 \leq t \leq 50$). Compared to 
Fig. 1(c) and Fig. 1(d) the effect of multi-frequency seems to be reduced in Fig. 2(c) and Fig. 2(d).

\begin{figure}[ht!]
\begin{center}
\includegraphics[height=5.0cm]{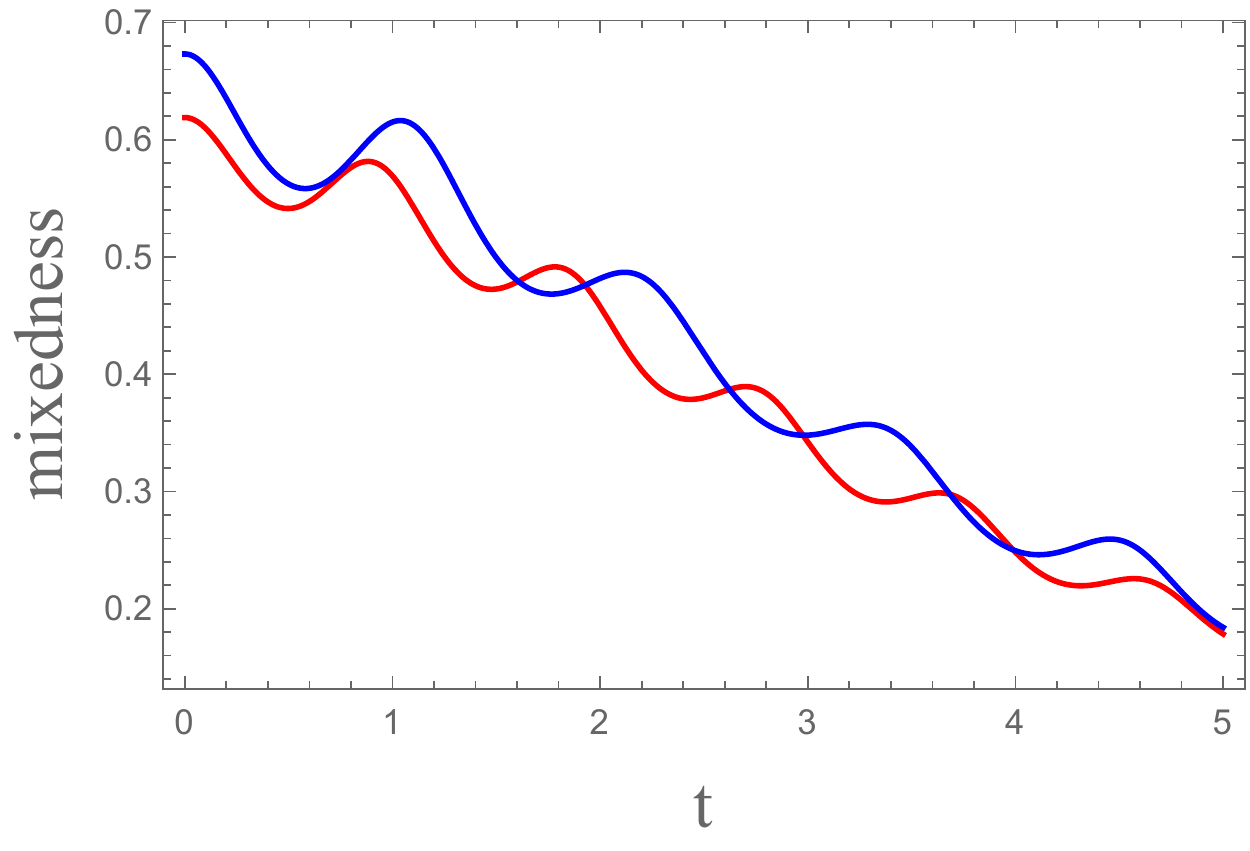} 
\includegraphics[height=5.0cm]{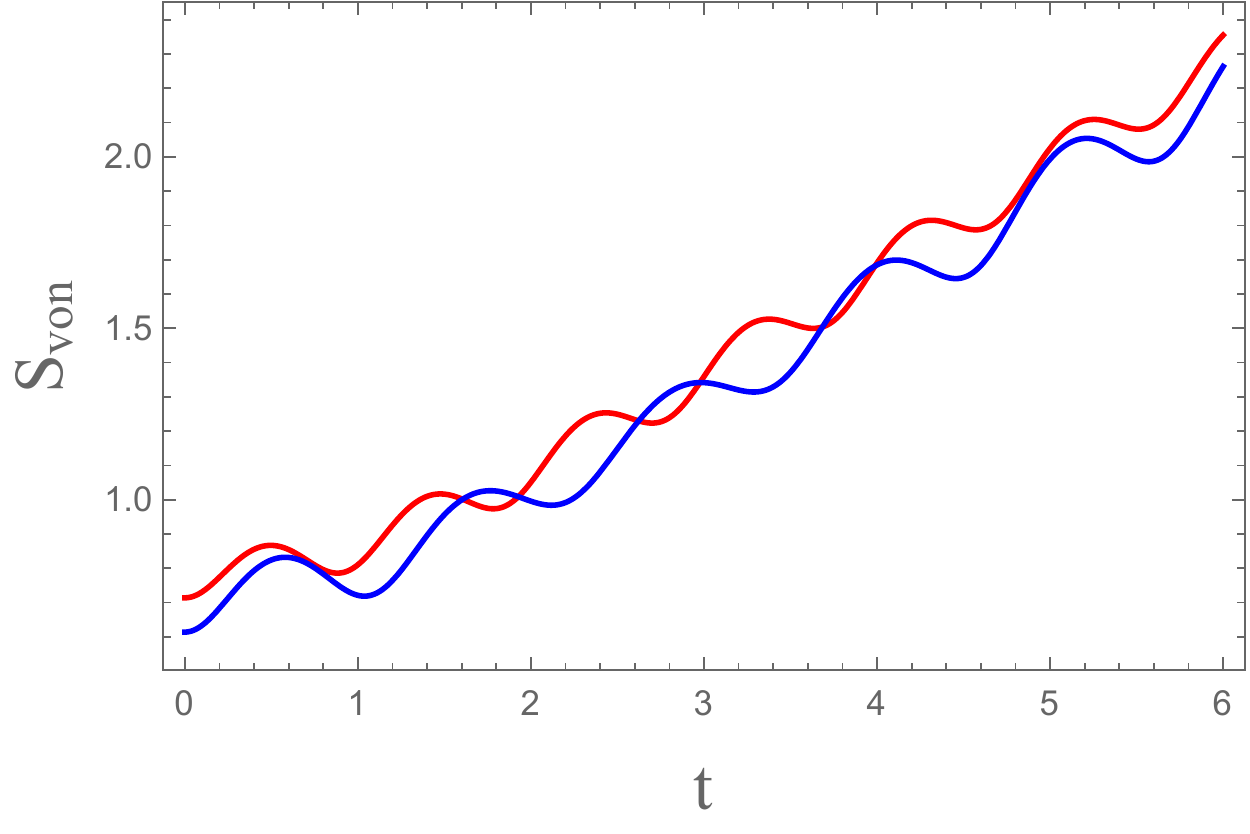}

\caption[fig3]{(Color online) The time-dependence of mixedness (Fig. 3(a)) and von Neumann entropy (Fig. 3(b)) when the quenched parameters are 
chosen as  $K_{0,i} = 0.1$,  $K_{0,f} = -0.1$,  $J_{12,i} = 1$, $J_{12,f} = 2$, $J_{13,i} = 2.5$, $J_{13,f} = 3.5$, $J_{23,i} = 3$, and $J_{23,f} = 4$.
The red and blue lines correspond to $\rho_C^{(red)}$ and $\rho_{BC}^{(red)}$ respectively. Since negative $K_{0,f}$ yields pure imaginary $\omega_{1,f}$,
the mixedness and von Neumann entropy exhibit exponential behavior with oscillation generated by $\omega_{+}$ and $\omega_{-}$.  }
\end{center}
\end{figure}
For completeness, finally, we examine the effect of negative frequency parameter although it is not physical situation. For this we choose
$K_{0,i} = 0.1$ and $K_{0,f} = -0.1$, which result in $\omega_{1,i} = 0.316$ and $\omega_{1,f} = 0,316 i$. The pure imaginary value of  $\omega_{1,f}$ 
changes  the cosine factor in $b_1 (t)$ into hyperbolic function. Thus, the dynamics of mixedness and entanglement should exhibit oscillatory and 
exponential behaviors. The remaining parameters are chosen as the same with second example. Then $\omega_{\pm}$ become
$\omega_{+,i} = 2.90$, $\omega_{-,i} = 2.19$, $\omega_{+,f} = 3,35$, and $\omega_{-,f} = 2.76$. In Fig. 3(a) the time-dependence of 
$\mbox{tr} \left[ \left(\rho_{BC}^{(red)} \right)^2 \right]$ (blue line) and $\mbox{tr} \left[ \left(\rho_{C}^{(red)} \right)^2 \right]$ (red line) is plotted.
As expected both exhibit exponential decay with oscillatory behavior. Like the previous models $\mbox{tr} \left[ \left(\rho_{BC}^{(red)} \right)^2 \right]$
is larger than  $\mbox{tr} \left[ \left(\rho_{C}^{(red)} \right)^2 \right]$ in most time intervals. In Fig. 3(b) the time-dependence of 
 $S_{von}^C$ (red line) and  $S_{von}^{BC}$ (blue line) is plotted. As expected both also exhibit exponential behavior with oscillation. The unexpected fact
 is the fact that the von Neumann entropies increase with increasing time. Usually completely mixed state has zero entanglement in the qubit system. Thus 
 we expect the decreasing behavior of the von Neumann entropies with increasing time. Fig. 3(b) shows an opposite behavior. Similar behavior can be 
 seen in the two coupled oscillator system with imaginary frequency (see Fig. 2(a) of Ref. \cite{park18}). Probably, this is mainly due to the fact that 
 this third example is unphysical because of negative frequency parameter.

\section{Conclusions}
The dynamics of mixedness and  entanglement is derived analytically by solving the TDSE of the three coupled harmonic oscillator system when the frequency
parameter $K_0$ and coupling constants $J_{ij}$ are arbitrarily time-dependent. For the calculation we assume that part of oscillator(s) is inaccessible. 
Thus we derive the dynamics of entanglement between inaccessible and accessible oscillators. To show the dynamics pictorially we 
introduce three sudden quenched models, where Ermakov equation (\ref{ermakov-1}) can be solved analytically. As expected due to the scale factors
$b_{j} (t)$, both mixedness and entanglement exhibit oscillatory behavior with multi-frequencies. It is shown that the mixedness for the case of one 
inaccessible oscillator is larger than that for the case of two inaccessible oscillators in the most time interval. Contrary to the mixedness entanglement for the
case of one inaccessible oscillator is smaller  than that  for the case of two inaccessible oscillators in the most time interval. 

It is natural to extend this paper to $n$-coupled harmonic oscillator system with arbitrary time-dependent frequency and coupling parameters, whose 
Hamiltonian can be written as 
\begin{equation}
\label{n-oscillator}
H = \frac{1}{2} \sum_{i=1}^n p_i^2 + \frac{1}{2} \left[ K_0 (t) \sum_{i=1}^n x_i^2 + \sum_{i < j}^n J_{ij} (t) (x_i - x_j)^2 \right].
\end{equation}
Generalizing the method presented in this paper we think the TDSE of this $n$-oscillator system can be solved analytically. Assuming that $m$-oscillator(s)
is inaccessible, it seems to be possible to derive the time-dependence of entanglement between inaccessible and accessible oscillators. It is of interest
to examine the effect of $m$ with fixed $n$ or effect of $n$ with fixed $m$ in the dynamics of entanglement. 

Another interesting issue related to this paper is how to compute the tripartite entanglement of the total state (\ref{density-1}). In qubit system it is possible 
to compute the three-tangle for any three-qubit pure state\cite{ckw}. However, this cannot be directly applied to our realistic system. Probably, we need new 
computable entanglement measure to explore this issue. We hope to visit this issue in the future.

{\bf Acknowledgement}:
This work was supported by the Kyungnam University Foundation Grant, 2018.

\newpage 

\begin{appendix}{\centerline{\bf Appendix A}}

\setcounter{equation}{0}
\renewcommand{\theequation}{A.\arabic{equation}}

The explicit expressions of quantities $\alpha_i$, $\beta_i$, and $\gamma_{ij}$ in Eq. (\ref{reduceBC-3}) are as follows:
\begin{eqnarray}
\label{alpha-1}
&&\alpha_1 = \frac{1}{36} |v_1|^2 + \frac{1}{4} |v_+|^2 A_+^4 (-J_{12} + J_{23} - z)^2 (J_{12} - J_{13} + z)^2      \\   \nonumber
&& \hspace{2.0cm} + \frac{1}{4} |v_-|^2 A_-^4 (-J_{12} + J_{23} + z)^2 (J_{12} - J_{13} - z)^2                           \\    \nonumber
&&+ \frac{A_+^2}{6} \left[ Z_+^2 \omega'_1 \omega'_+ + \left(\omega'_1 \omega'_+ + \frac{\dot{b_1}}{b_1} \frac{\dot{b_+}}{b_+} \right)
 (J_{12} - J_{13} + z)  (-J_{12} + J_{23} - z) \right]                                                                                            \\   \nonumber
&&+ \frac{A_-^2}{6} \left[ Z_-^2 \omega'_1 \omega'_- + \left(\omega'_1 \omega'_- + \frac{\dot{b_1}}{b_1} \frac{\dot{b_-}}{b_-} \right)
 (J_{12} - J_{13} - z)  (-J_{12} + J_{23} + z) \right]                                                                                            \\    \nonumber
&&+ \frac{A_+^2 A_-^2}{2} \bigg[ 4 z^2 (J_{13} - J_{23})^2 \omega'_+ \omega'_-                                                 
+  \left(\omega'_+ \omega'_- + \frac{\dot{b_+}}{b_+} \frac{\dot{b_-}}{b_-} \right)  (J_{12} - J_{13} + z)        \\   \nonumber
&& \hspace{4.0cm} \times (J_{12} - J_{13} - z)   (-J_{12} + J_{23} + z)  (-J_{12} + J_{23} - z) \bigg]
 \end{eqnarray}
 
 \begin{eqnarray}
 \label{beta-1}
 && \beta_1 = \frac{A_+^2}{6} Z_+ \left[ \omega'_1  \frac{\dot{b_+}}{b_+}  (J_{12} - J_{13} + z) -  \frac{\dot{b_1}}{b_1} \omega'_+ 
 (-J_{12} + J_{23} - z)   \right]                                                                                                                            \\    \nonumber
 && \hspace{1.0cm} + \frac{A_-^2}{6} Z_- \left[ \omega'_1  \frac{\dot{b_-}}{b_-}  (J_{12} - J_{13} - z) -  \frac{\dot{b_1}}{b_1} \omega'_- 
 (-J_{12} + J_{23} + z)   \right]                                                                                                                            \\    \nonumber
 &&\hspace{1.0cm} +A_+^2 A_-^2 z  (J_{13} - J_{23}) \bigg[ \omega'_+  \frac{\dot{b_-}}{b_-}  (J_{12} - J_{13} - z)  (-J_{12} + J_{23} - z)
                                                                                                                                                                          \\    \nonumber
&& \hspace{5.0cm} - \frac{\dot{b_+}}{b_+} \omega'_-  (J_{12} - J_{13} + z)  (-J_{12} + J_{23} + z)  \bigg]
 \end{eqnarray}

\begin{eqnarray}
\label{alpha-2}
&&\alpha_2 = \frac{1}{36} |v_1|^2 + \frac{1}{4} |v_+|^2 A_+^4  (J_{13} - J_{23})^2 (-J_{12} + J_{23} - z)^2       \\     \nonumber 
&& \hspace{4.0cm} + \frac{1}{4} |v_-|^2 A_-^4  (J_{13} - J_{23})^2 (-J_{12} + J_{23} + z)^2                           \\    \nonumber
&&+ \frac{A_+^2}{6} \left[ (J_{12} + J_{13} - 2 J_{23} + z)^2 \omega'_1 \omega'_+ + \left(\omega'_1 \omega'_+ + \frac{\dot{b_1}}{b_1} \frac{\dot{b_+}}{b_+} \right)  (J_{13} - J_{23})  (-J_{12} + J_{23} - z) \right]                                                     \\   \nonumber
&& + \frac{A_-^2}{6} \left[ (J_{12} + J_{13} - 2 J_{23} - z)^2 \omega'_1 \omega'_- + \left(\omega'_1 \omega'_- + \frac{\dot{b_1}}{b_1} \frac{\dot{b_-}}{b_-} \right)  (J_{13} - J_{23})  (-J_{12} + J_{23} + z) \right]                                                       \\    \nonumber
&&+ \frac{A_+^2 A_-^2}{2}  (J_{13} - J_{23})^2 \left[ 4 z^2 \omega'_+ \omega'_-                                                 
+  \left(\omega'_+ \omega'_- + \frac{\dot{b_+}}{b_+} \frac{\dot{b_-}}{b_-} \right)   (-J_{12} + J_{23} + z)  (-J_{12} + J_{23} - z) \right]
 \end{eqnarray}

 \begin{eqnarray}
 \label{beta-2}
 && \beta_2 = \frac{A_+^2}{6}  (J_{12} + J_{13} - 2 J_{23} + z) \left[ \omega'_1  \frac{\dot{b_+}}{b_+}  (J_{13} - J_{23}) -  \frac{\dot{b_1}}{b_1} \omega'_+ (-J_{12} + J_{23} - z)   \right]                                                                                                                            \\    \nonumber
 && \hspace{.8cm}   + \frac{A_-^2}{6}  (J_{12} + J_{13} - 2 J_{23} - z) \left[ \omega'_1  \frac{\dot{b_-}}{b_-}  (J_{13} - J_{23}) -  \frac{\dot{b_1}}{b_1} \omega'_-  (-J_{12} + J_{23} + z)   \right]                                                                                                                   \\    \nonumber
 && \hspace{.8cm}- A_+^2 A_-^2 z  (J_{13} - J_{23})^2 \bigg[ \omega'_+  \frac{\dot{b_-}}{b_-}   (-J_{12} + J_{23} - z)
                                  - \frac{\dot{b_+}}{b_+} \omega'_-   (-J_{12} + J_{23} + z)  \bigg]
 \end{eqnarray}

 \begin{eqnarray}
\label{alpha-3}
&&\alpha_3 = \frac{1}{36} |v_1|^2 + \frac{1}{4} |v_+|^2 A_+^4  (J_{13} - J_{23})  (J_{12} - J_{13} + z) (-J_{12} + J_{23} - z)^2       \\     \nonumber 
&& \hspace{3.0cm} + \frac{1}{4} |v_-|^2 A_-^4  (J_{13} - J_{23}) (J_{12} - J_{13} - z) (-J_{12} + J_{23} + z)^2                           \\    \nonumber
&&+ \frac{A_+^2}{12} \left[2 Z_+ (J_{12} + J_{13} - 2 J_{23} + z) \omega'_1 \omega'_+ - \left(\omega'_1 \omega'_+ + \frac{\dot{b_1}}{b_1} \frac{\dot{b_+}}{b_+} \right)   (-J_{12} + J_{23} - z)^2 \right]                                                     \\   \nonumber
&&  + \frac{A_-^2}{12} \left[2 Z_- (J_{12} + J_{13} - 2 J_{23} - z) \omega'_1 \omega'_- - \left(\omega'_1 \omega'_- + \frac{\dot{b_1}}{b_1} \frac{\dot{b_-}}{b_-} \right)   (-J_{12} + J_{23} + z)^2 \right]                                                   \\    \nonumber
&&+ \frac{A_+^2 A_-^2}{2}  (J_{13} - J_{23}) \bigg[ -4 z^2  (J_{13} - J_{23}) \omega'_+ \omega'_-                                                 
+  \left(\omega'_+ \omega'_- + \frac{\dot{b_+}}{b_+} \frac{\dot{b_-}}{b_-} \right) (J_{12} - J_{13})         \\     \nonumber
&& \hspace{6.0cm} \times
  (-J_{12} + J_{23} + z)  (-J_{12} + J_{23} - z) \bigg]
 \end{eqnarray}
 
 \begin{eqnarray}
 \label{beta-3}
 && \beta_3 = \frac{A_+^2}{12} \bigg[ \omega'_1  \frac{\dot{b_+}}{b_+} 
\left\{ 2 (J_{13} - J_{23})  (J_{12} - J_{13} + z) +  (-J_{12} + J_{23} - z)^2 \right\}                                         \\   \nonumber
&& \hspace{6.0cm}+ 3 \frac{\dot{b_1}}{b_1} \omega'_+ (-J_{12} + J_{23} - z)^2   \bigg]                              \\    \nonumber
&& \hspace{1.0cm}+ \frac{A_-^2}{12} \bigg[ \omega'_1  \frac{\dot{b_-}}{b_-} 
\left\{ 2 (J_{13} - J_{23})  (J_{12} - J_{13} - z) +  (-J_{12} + J_{23} + z)^2 \right\}                                         \\   \nonumber
&& \hspace{6.0cm}+ 3 \frac{\dot{b_1}}{b_1} \omega'_-   (-J_{12} + J_{23} + z)^2   \bigg]                              \\    \nonumber                                                      
&&\hspace{.8cm} +\frac{A_+^2 A_-^2}{2} z  (J_{13} - J_{23}) \bigg[ - \omega'_+  \frac{\dot{b_-}}{b_-}   (-J_{12} + J_{23} - z)  (J_{12} - 2J_{13} + J_{23} - z)
                                                                                                                                                                        \\    \nonumber
&&\hspace{3.0cm}
                                  + \frac{\dot{b_+}}{b_+} \omega'_-   (-J_{12} + J_{23} + z)  (J_{12} - 2J_{13} + J_{23} + z) \bigg]
\end{eqnarray}
 
 \begin{eqnarray}
\label{alpha-4}
&&\alpha_4 = \frac{1}{36} |v_1|^2 + \frac{1}{4} |v_+|^2 A_+^4  (J_{13} - J_{23})  (J_{12} - J_{13} + z) (-J_{12} + J_{23} - z)^2       \\     \nonumber 
&& \hspace{3.0cm} + \frac{1}{4} |v_-|^2 A_-^4  (J_{13} - J_{23}) (J_{12} - J_{13} - z) (-J_{12} + J_{23} + z)^2                           \\    \nonumber
&&- \frac{A_+^2}{12}    (-J_{12} + J_{23} - z)^2     \left(\omega'_1 \omega'_+ + \frac{\dot{b_1}}{b_1} \frac{\dot{b_+}}{b_+} \right)                                          
- \frac{A_-^2}{12}    (-J_{12} + J_{23} + z)^2     \left(\omega'_1 \omega'_- + \frac{\dot{b_1}}{b_1} \frac{\dot{b_-}}{b_-} \right)                                                                                                                  
                                                                                                                                                                                     \\    \nonumber
&&+ \frac{A_+^2 A_-^2}{2}   (J_{12} - J_{13})  (J_{13} - J_{23})  (-J_{12} + J_{23} + z)  (-J_{12} + J_{23} - z)
\left(\omega'_+ \omega'_- + \frac{\dot{b_+}}{b_+} \frac{\dot{b_-}}{b_-} \right)
 \end{eqnarray}
 
 \begin{eqnarray}
 \label{beta-4}
 && \beta_4 = \frac{A_+^2}{12}  (-J_{12} + J_{23} - z) (J_{12} - 2 J_{13} + J_{23} + z)                                                                                                                             \left( \omega'_1 \frac{\dot{b_+}}{b_+} -   \frac{\dot{b_1}}{b_1} \omega'_+ \right)             \\    \nonumber
 && \hspace{1.0cm} +  \frac{A_-^2}{12}  (-J_{12} + J_{23} + z) (J_{12} - 2 J_{13} + J_{23} - z)                                                                                                                             \left( \omega'_1 \frac{\dot{b_-}}{b_-} -   \frac{\dot{b_1}}{b_1} \omega'_- \right)                                                                                                               \\    \nonumber
 && \hspace{.8cm} -\frac{A_+^2 A_-^2}{2}  z  (J_{13} - J_{23})  (-J_{12} + J_{23} + z)  (-J_{12} + J_{23} - z)
 \left( \omega'_+ \frac{\dot{b_-}}{b_-} -   \frac{\dot{b_+}}{b_+} \omega'_- \right)                                                                                                          
 \end{eqnarray}

\begin{eqnarray}
\label{gamma-11}
&&\gamma_{11} = \frac{1}{36} |v_1|^2 + \frac{1}{4} |v_+|^2 A_+^4  (J_{12} - J_{13} + z)^2 (-J_{12} + J_{23} - z)^2      \\   \nonumber   
&& \hspace{4.0cm} + \frac{1}{4} |v_-|^2 A_-^4  (J_{12} - J_{13} - z)^2 (-J_{12} + J_{23} + z)^2                           \\    \nonumber
&&\hspace{3.0cm}+ \frac{A_+^2}{12} (v_1 v_+^* + v_1^* v_+)   (J_{12} - J_{13} + z)  (-J_{12} + J_{23} - z)                                     \\     \nonumber
&&\hspace{3.0cm}+ \frac{A_-^2}{12} (v_1 v_-^* + v_1^* v_-)   (J_{12} - J_{13} - z)  (-J_{12} + J_{23} + z)                                    \\    \nonumber
&&\hspace{.5cm}+ \frac{A_+^2 A_-^2}{4}   (v_+ v_-^* + v_+^* v_-)  (J_{12} - J_{13} + z)  (J_{12} - J_{13} - z)  (-J_{12} + J_{23} + z)   (-J_{12} + J_{23} - z)  
\end{eqnarray}
 
\begin{eqnarray}
\label{gamma-22}
&&\gamma_{22} = \frac{1}{36} |v_1|^2 + \frac{1}{4} |v_+|^2 A_+^4  (J_{13} - J_{23})^2 (-J_{12} + J_{23} - z)^2       \\   \nonumber
&& \hspace{4.0cm}+ \frac{1}{4} |v_-|^2 A_-^4  (J_{13} - J_{23})^2 (-J_{12} + J_{23} + z)^2                           \\    \nonumber
&&\hspace{3.0cm}+ \frac{A_+^2}{12} (v_1 v_+^* + v_1^* v_+)  (J_{13} - J_{23})  (-J_{12} + J_{23} - z)                                     \\     \nonumber
&&\hspace{3.0cm}+ \frac{A_-^2}{12} (v_1 v_-^* + v_1^* v_-)   (J_{13} - J_{23}) (-J_{12} + J_{23} + z)                                    \\    \nonumber
&&\hspace{.5cm}+ \frac{A_+^2 A_-^2}{4}   (v_+ v_-^* + v_+^* v_-)  (J_{13} - J_{23})^2  (-J_{12} + J_{23} + z)   (-J_{12} + J_{23} - z).  
\end{eqnarray}
 
 \end{appendix}
 
 \newpage 

\begin{appendix}{\centerline{\bf Appendix B}}

\setcounter{equation}{0}
\renewcommand{\theequation}{B.\arabic{equation}}
Since $\rho_{ABC}$ is pure state, it is easy to show that the R\'{e}nyi and von Neumann entropies of $\rho_{BC}^{(red)}$ are exactly the same with those 
of $\rho_A^{(red)}$. Thus, we can compute the entropies of $\rho_{BC}^{(red)}$ by solving 
\begin{equation}
\label{eigenA-1}
\int dy  \rho_{A}^{(red)} (x, y : t) g_{n} (y : t) = q_{n} (t) g_{n} (x : t).
\end{equation}

It is straightforward to show that the explicit expression of $\rho_{A}^{(red)}$ is 
\begin{eqnarray}
\label{reduceA-1}
&&\rho_A{(red)} (x, y : t)                                  \\    \nonumber
&=&  \int dx_2 dx_3 \rho_{ABC} (x, x_2, x_3 : y, x_2, x_3 : t)                            \\     \nonumber
&=& \left( \frac{\omega'_1 \omega'_+ \omega'_-}{\pi \Omega_A} \right)^{1/2} \exp \left[ - \frac{1}{\Omega_A} 
\left\{ (R_A - i I_A) x^2 + (R_A + i I_A) y^2 - 2 Y_A x y \right\}  \right]
\end{eqnarray}
where
\begin{eqnarray}
\label{reduceA-2}
&&\Omega_A = \frac{1}{3} \left[A_+^2 X_+^2 \omega'_1 \omega'_+ + A_-^2 X_-^2 \omega'_1 \omega'_- + \omega'_+ \omega'_- \right]
                                                                                                                                                                                          \\    \nonumber
&& Y_A = \frac{|v_1|^2}{36} \left( A_+^2 X_+^2  \omega'_+ + A_-^2 X_-^2  \omega'_- \right)                                                 \\    \nonumber
&&\hspace{1.0cm} + \frac{\omega'_1}{12} \left[ A_+^4 X_+^2 (-J_{12} + J_{23} - z)^2 |v_+|^2 + A_-^4 X_-^2 (-J_{12} + J_{23} + z)^2 |v_-|^2\right]
                                                                                                                                                                                           \\   \nonumber
&&\hspace{1.0cm} + A_+^2 A_-^2 z^2 (J_{13} - J_{23})^2 \left[ A_+^2 (-J_{12} + J_{23} - z)^2 |v_+|^2 \omega'_- + A_-^2 (-J_{12} + J_{23} + z)^2 \omega'_+ |v_-|^2 \right]
                                                                                                                                                                                           \\    \nonumber
&&\hspace{1.0cm}+ \frac{A_+^2 A_-^2}{4} (J_{13} - J_{23}) \Bigg[ -(J_{12} - J_{13}) (-J_{12} + J_{23} + z) (-J_{12} + J_{23} - z) \omega'_1 (v_+ v_-^* + v_+^* v_-)
                                                                                                                                                                                            \\    \nonumber
&&\hspace{1.0cm}+ \frac{2}{3} z X_+ (-J_{12} + J_{23} + z) \omega'_+ (v_1 v_-^* + v_1^* v_-) - \frac{2}{3} z X_- (-J_{12} + J_{23} - z) \omega'_- (v_1 v_+^* + v_1^* v_+)
                                                                                                                                        \Bigg]                                       \\   \nonumber
&&R_A = Y_A + \frac{1}{2} \omega'_1 \omega'_+ \omega_-                                                                                               \\   \nonumber
&& I_A =- A_+^2 A_-^2 z (J_{13} - J_{23}) \bigg[ X_+ (-J_{12} + J_{23} + z) \omega'_1 \omega'_+ \frac{\dot{b}_-}{b_-} 
- X_- (-J_{12} + J_{23} - z) \omega'_1 \frac{\dot{b}_+}{b_+} \omega'_-                                                                            \\   \nonumber
&&\hspace{8cm}- 2 z (J_{13} - J_{23}) \frac{\dot{b}_1}{b_1} \omega'_+ \omega'_- \bigg]
\end{eqnarray}
with $X_{\pm} = J_{12} + J_{23} - 2 J_{13} \pm z$. It is useful to note $X_+ X_- = -3 (J_{12} - J_{13}) (J_{13} - J_{23})$ and  
\begin{equation}
\label{reduceA-3}
X_{\pm} (-J_{12} + J_{23} \mp z) = - (J_{12} - J_{13}) Z_{\pm} \hspace{.5cm} 
X_{\pm} (-J_{12} + J_{23} \pm z ) = - (J_{13} - J_{23}) W_{\pm}
\end{equation}
where $Z_{\pm} = 2 J_{12} - J_{13} - J_{23} \pm 2 z$ and $W_{\pm} = 2 J_{23} - J_{12} - J_{13} \pm 2 z$. 

Following the case of $\rho_C^{(red)}$ it is straightforward to show that the eigenvalue of Eq. (\ref{eigenA-1}) is $q_n (t) = (1 - \xi_A) \xi_A^n$ and, 
the R\'{e}nyi and von Neumann entropies of $\rho_{BC}^{(red)}$ are given by 
 \begin{eqnarray}
 \label{eigenBC-f}
 &&S_{\alpha}^{BC} \equiv \frac{1}{1 - \alpha} \ln \mbox{tr} \left[ \left(\rho_{BC}^{(red)} \right)^{\alpha} \right] = \frac{1}{1 - \alpha} 
 \ln \frac{(1 - \xi_A)^{\alpha}}{1 - \xi_A^{\alpha}}                                                                                     \\    \nonumber
 &&S_{von}^{BC} = \lim_{\alpha \rightarrow 1 } S_{\alpha}^{BC} = -\ln (1 - \xi_A) - \frac{\xi_A}{1 - \xi_A} \ln \xi_A
 \end{eqnarray}
where 
\begin{equation}
\xi_A = \frac{Y_A}{R_A + \sqrt{R_A^2 - Y_A^2}}.
\end{equation}

Although we have not proved analytically that Eq. (\ref{eigenBC-f}) and Eq. (\ref{eigenBC-24}) are exactly the same due to long expressions
introduced in appendix A, this coincidence  is confirmed numerically when plotting Fig. 1(d), Fig. 2(d), and Fig. 3(b).

\end{appendix}

\end{document}